\documentclass[a4paper,10pt]{article}

\usepackage{jheppub} 

\usepackage[T1]{fontenc} 
\usepackage{lmodern} 
\usepackage[toc,page]{appendix}

\usepackage{xcolor}
\usepackage{amsmath,amssymb}
\usepackage{mathrsfs,wasysym}
\usepackage{booktabs}
\usepackage{slashed}
\usepackage{cancel}
\usepackage{bm}
\usepackage{physics}
\usepackage{placeins}
\usepackage{BOONDOX-cal}

\newcommand{\xl}{\bar x}
\newcommand{\yl}{\bar y}
\newcommand{\zl}{\bar z}
\newcommand{\el}{\bar\eta}
\newcommand{\kvec}{\textbf{k}}
\newcommand{\pvec}{\textbf{p}}
\newcommand{\qvec}{\textbf{q}}
\newcommand{\Dvec}{\bm{\Delta}}
\newcommand{\qt}{q_{\rm t}}
\newcommand{\pt}{p_{\rm t}}
\newcommand{\kt}{k_{\rm t}}

\newcommand{\qthat}{\hat q_{\rm t}}
\newcommand{\pthat}{\hat p_{\rm t}}
\newcommand{\MSbar}{\overline{\text{MS}}}

\newcommand{\sss}{\scriptscriptstyle}
\newcommand{\as}{\alpha_s}
\newcommand{\Ord}{\mathcal{O}}
\newcommand{\Lum}{\mathscr{L}}

\newcommand{\muf}{\mu_{\sss\rm F}}
\newcommand{\mur}{\mu_{\sss\rm R}}

\newcommand{\plus}[1]{\left(#1\right)_+}

\let\originalleft\left
\let\originalright\right
\renewcommand{\left}{\mathopen{}\mathclose\bgroup\originalleft}
\renewcommand{\right}{\aftergroup\egroup\originalright}

\newcommand{\dvec}[2]{\left(
    \begin{array}{c}
      #1 \\ #2
    \end{array}
\right)}

\newcommand{\pdfcoll}[2]{f_{#1}\qty(#2)}

\newcommand{\myexp}[1]{\mathrm{e}^{#1}}

\newcommand{\thetapair}{\psi}
\newcommand{\thetap}{\vartheta'}

\def\beq{\begin{equation}}  
\def\eeq{\end{equation}}
\def\({\left(}
\def\){\right)}
\def\[{\left[}
\def\]{\right]}

\let\oldsubsection\subsection
\renewcommand\subsection[2][\subsectiontoc]{%
  \def\subsectiontoc{#2}%
  \oldsubsection[#1]{\boldmath #2}%
}

\let\oldsubsubsection\subsubsection
\renewcommand\subsubsection[2][\subsubsectiontoc]{%
  \def\subsubsectiontoc{#2}%
  \oldsubsubsection[#1]{\boldmath #2}%
}

\newcommand{\hell}{\href{https://www.roma1.infn.it/~bonvini/hell/}{\texttt{HELL}}}

\allowdisplaybreaks

\title{\boldmath Differential heavy quark pair production at small $x$}

\author[a,b,c]{Federico Silvetti}
\author[b]{and Marco Bonvini}
\affiliation[a]{Dipartimento di Fisica, Sapienza Universit\`a di Roma,\\ Piazzale Aldo Moro~5, 00185 Roma, Italy}
\affiliation[b]{INFN, Sezione di Roma 1,\\ Piazzale Aldo Moro~5, 00185 Roma, Italy}
\affiliation[c]{Theoretical Physics Department, CERN,\\ 1211 Geneva 23, Switzerland}
\preprint{CERN-TH-2022-186}

\emailAdd{federico.silvetti@uniroma1.it}
\emailAdd{marco.bonvini@roma1.infn.it}

\abstract{%
We consider the production of a heavy quark pair in proton-proton collisions.
For bottom and charm quarks, the final state invariant mass is typically much smaller
than the collider energy (e.g.\ at the LHC), so that high-energy logarithms
may spoil the perturbativity of the theoretical prediction at fixed order.
The resummation of these logarithms to all orders is thus needed to obtain reliable predictions.
In this work, we extend previous results on high-energy (or small-$x$) resummation
to differential distributions in rapidity, transverse momentum and invariant mass,
and implement them in the public code \texttt{HELL}.
}

\begin{document}

\maketitle


\section{Introduction}

In the era of LHC precision physics, considerable efforts are required
to match theoretical prediction with experimental accuracy.
Such an endeavour requires several different inputs, e.g.\ high-order predictions for partonic processes,
high-quality parton distributions and all-order resummation of large logarithmic contributions.

In this work, we focus on the latter and specifically on the so-called high-energy logarithms of the form
$\as^n\frac1x\log^k\frac1x$, $k<n$, where $x$ is a dimensionless scaling variable that becomes small when the collider energy $s$ is large.
These perturbative terms arise beyond the leading order in both the partonic cross sections
and the DGLAP splitting functions governing PDF evolution (in $\MSbar$-like schemes).
At the energy scales of many LHC processes, $x \ll 1$ and these logarithms spoil the perturbativity of the fixed-order results.
This calls for an all-order resummation of these corrections.

The theoretical framework to perform this high-energy (or small-$x$) resummation has been established during the last thirty years
starting with the resummation of splitting functions~\cite{Salam:1998tj,Ciafaloni:1999yw,Ciafaloni:2003kd,Ciafaloni:2003rd,Ciafaloni:2007gf,Ball:1995vc,Ball:1997vf,Altarelli:2001ji,Altarelli:2003hk,Altarelli:2005ni,Altarelli:2008aj,Thorne:1999sg,Thorne:1999rb,Thorne:2001nr,White:2006yh,Rothstein:2016bsq}
by means of the Balitsky-Fadin-Kuraev-Lipatov (BFKL) equation~\cite
{Lipatov:1976zz,Fadin:1975cb,Kuraev:1976ge,Kuraev:1977fs,Balitsky:1978ic,Fadin:1998py}
and arriving recently to PDF determination with resummed theory~\cite{Ball:2017otu,Abdolmaleki:2018jln,Bonvini:2019wxf}.

One of the key steps to achieve a consistent resummed prediction is the resummation of partonic cross sections,
which can be carried out to leading logarithmic (LL) precision using the $\kt$ factorization theorem~\cite
{Catani:1990xk,Catani:1990eg,Catani:1994sq}.
Recently, the resummation technique for partonic cross sections has been reformulated
and adapted for stable numerical implementation~\cite{Bonvini:2016wki,Bonvini:2017ogt,Bonvini:2018iwt}.
This led to the release of the High-Energy Large Logarithms (\hell) public code,
which aims to provide a systematic framework for implementing small-$x$ resummation.

So far, only inclusive observables have been considered in \hell.
The sensitivity of inclusive observables to resummation effects is, however, limited.
Indeed, the small-$x$ region at parton level is mixed with the medium- and high-$x$ regions in the
convolution that defines the hadron-level cross section, thereby smoothening out
much of the impact of high-energy logarithms (see e.g.~Ref.~\cite{Bonvini:2018iwt}).
Differential distributions, instead, can be more directly sensitive to
specific values of partonic $x$,
thereby enhancing the effect of small-$x$ resummation in some kinematic regions.
Moreover differential distributions are of greater phenomenological interest,
as they can be compared more directly with experimental measurements.

In this work we will focus on invariant mass, rapidity and transverse momentum distributions.
The resummation of small-$x$ logarithms in these differential cross sections
was developed in Refs.~\cite{Caola:2010kv,Forte:2015gve,Muselli:2017fuf}, focussing on Higgs production via gluon fusion.
Here, we revisit these results and extend them to the modern resummation formalism of
Refs.~\cite{Bonvini:2016wki,Bonvini:2017ogt,Bonvini:2018iwt},
thereby allowing for a stable numerical implementation thus opening the door to phenomenological studies.

We apply our findings to heavy flavour pair production,
and construct resummed predictions for distribution in invariant mass, rapidity and transverse momentum
of either the heavy-quark pair or one of the heavy quarks.
This process is particularly interesting due to the availability of measurements from the LHCb experiment
for the production of charm and bottom quarks in the forward region,
where one of the incoming partons is certainly at small $x$ and thus the effect of resummation should be marked.
In addition, these data reach values of $x$ down to $x \sim {10}^{-6}$,
which is a region of proton momentum fractions so far unexplored,
as the HERA data is limited to $x \gtrsim 3\cdot10^{-5}$ in the perturbative regime.
Our results thus provide an important ingredient to refine the determination of PDFs at small-$x$,
which serves both as a test of QCD in extreme regimes and as a tool to improve
high-energy phenomenology.
All our results are available through the new release of the \hell\ code.

The structure of this paper is the following.
Section \ref{sec:fllydiff} is dedicated to presenting the formalism of $\kt$ factorization in a proton collider
and its use to construct small-$x$ resummed results for differential distributions in the language of \hell.
Then, section \ref{sec:HQ} is dedicated to the application of resummation to differential heavy flavour production,
parametrising the final state respectively as the entire quark-antiquark pair or as a single quark.
We conclude in section \ref{sec:conclusions}, and collect in the appendices various details on
analytical expressions for heavy quark production and aspects of numerical implementation.

\section{Multi-differential small-$x$ resummation in \texttt{HELL}}
\label{sec:fllydiff}

The resummation of small-$x$ logarithms in physical observables is based on
$\kt$ factorization~\cite{Catani:1990xk,Catani:1990eg,Collins:1991ty,Catani:1993ww,Catani:1993rn,Catani:1994sq}.
The basic observation is that the leading small-$x$ logarithms arise,
in a physical gauge, from $\kt$ integration over gluon exchanges in the $t$ channel.
Therefore, in the small-$x$ limit, the generic amplitude squared can be decomposed into
contributions that are two-gluon irreducible (2GI) in the $t$ channel and thus do not contain any logarithmic enhancement.
Instead, the small-$x$ logarithms are produced by the integration over the momenta of the gluons connecting these 2GI block.
In this way the cross section of the process factorizes~\cite{Catani:1990xk,Catani:1990eg,Catani:1994sq}
into a process dependent 2GI coefficient, called off-shell coefficient function,
and process independent ``unintegrated'' PDFs that contain the traditional collinear PDFs and the sum over all possible
process independent 2GI kernels connected by off-shell gluons.
By making explicit the dependence of unintegrated PDF on collinear PDFs and comparing the result with
the standard collinear factorization, one finally obtains an expression for the LL resummation
of small-$x$ logarithms in the collinear partonic coefficient functions.

The last step of this procedure was traditionally performed in Mellin moment space,
which allows to obtain rather simple analytical expressions.
Despite the elegance of this result,
it was soon realized that subleading effects due to the running of the strong coupling
are important and should be included systematically in the resummation procedure to obtain
perturbatively stable results~\cite{Ball:2007ra,Altarelli:2008aj}.
However, the inclusion of such terms in Mellin space is complicated, and not suitable for efficient numerical implementations.
Recently, an alternative but equivalent formulation of the resummation was proposed~\cite{Bonvini:2016wki},
that solves the technical limitations of the original formulation by working directly in $\kt$ space,
leading to an efficient numerical implementation.
This novel approach is at the core of the public code \hell,
and allowed for a number of phenomenological applications~\cite{Bonvini:2017ogt,Bonvini:2018iwt},
including the first consistent PDF fits with small-$x$ resummation~\cite{Ball:2017otu,Abdolmaleki:2018jln,Bonvini:2019wxf}.

So far, all \hell\ applications are for inclusive observables (DIS structure functions~\cite{Bonvini:2016wki,Bonvini:2017ogt}
and the total Higgs production cross section~\cite{Bonvini:2018iwt}).
The resummation of differential observables, of obvious interest for LHC phenomenology,
has been considered in the Mellin-space formalism.
Specifically, resummed expressions for rapidity distributions~\cite{Caola:2010kv},
transverse momentum distributions~\cite{Forte:2015gve}
and double differential distributions in both rapidity and transverse momentum~\cite{Muselli:2017fuf}
are available.
It is the purpose of this section to reformulate these results in the new \hell\ language,
thereby supplementing them with the running coupling contributions and thus providing
a ready-to-use numerical implementation.

In this work, we focus on processes at hadron-hadron colliders that are gluon-gluon initiated at lowest order.
These include, for instance, Higgs production, jet production, or heavy quark pair production;
the latter will be considered as a practical application in section~\ref{sec:HQ}.
The reason for this choice is that the resummation is simpler, because at LL there are no collinear singularities.
In other processes where the lowest order is initiated by (massless) quarks, because small-$x$ logarithms at LL
appear from chains of emissions ending with a gluon, the diagram entering the computation of the off-shell coefficient function
must contain at least a gluon to (massless) quark splitting, thus producing a collinear singularity.
One example is the Drell-Yan process. In such cases, the collinear singularities must be treated at the resummed level
(similarly to what is done in DIS, see Ref.~\cite{Bonvini:2017ogt}).
A study of the Drell-Yan process where this issue is addressed at differential level is left to future work~\cite{DYsx}.

Before moving to the resummation, we establish the notation by presenting the structure of
differential distributions in collinear factorization for a process in proton-proton collisions.
We consider a generic final-state momentum $q$
(it can be the momentum of a single particle or the sum of momenta of different particles)
in the collider center-of-mass frame,
and we write the distribution differential in its invariant mass squared $Q^2\equiv q^2$,
rapidity $Y=\frac12\log\frac{q^0+q^3}{q^0-q^3}$ and transverse component squared $\qt^2=(q^1)^2+(q^2)^2$ as
\begin{align}\label{eq:collfactdiff}
\frac{\dd\sigma}{\dd Q^2\dd Y\dd \qt^2} \qty(\tau, Q^2,Y,\qt^2) &= \tau \sum_{ij} \int_\tau^1 \frac{\dd{x}}x\int \dd y\,
\frac{\dd C_{ij}}{\dd Q^2\dd y\dd \qt^2} \qty(x, Q^2,y,\qt^2,\as, \frac{Q^2}{\muf^2})
L_{ij}\qty(\frac\tau x, Y-y, \muf^2),
\end{align}
with $\tau=Q^2/s$ ($s$ is the collider energy squared) and the sum extends over all possible partons $i,j$ in each proton.
In this expression the function
\beq
\frac{\dd C_{ij}}{\dd Q^2\dd y\dd \qt^2} \qty(x, Q^2,y,\qt^2,\as, \frac{Q^2}{\muf^2})
\eeq
is the parton-level coefficient function,
which depends on $x=Q^2/\hat s_c$ (the parton-level analog of $\tau$)
where $\hat s_c$ is the partonic center-of-mass energy,\footnote
{We call it $\hat s_c$ ($c$ stands for collinear) because we will use $\hat s$ for the energy squared of another system.}
and on $y$ which is the rapidity of $q$ with respect to the partonic center-of-mass frame,
and is related to the proton-level rapidity $Y$ by a longitudinal boost.
Indeed, $Y-y$ is the rapidity of the partonic center-of-mass frame with respect to the collider frame,
and it is determined by the momentum fractions $\hat x_1,\hat x_2$
of the partons in each proton by $Y-y=\frac12\log\frac{\hat x_1}{\hat x_2}$.
Note that we have omitted the dependence of $\as$ and of the coefficient function on the renormalization scale $\mur$,
as such dependence is subleading in the small-$x$ limit we are interested in.
Finally, the function
\begin{equation}\label{eq:lumi}
  L_{ij}\qty(\xl, \yl, \muf^2) = \pdfcoll{i}{\sqrt{\xl}\myexp{\yl}, \muf^2} \pdfcoll{j}{\sqrt{\xl}\myexp{-\yl}, \muf^2} \theta\qty(\myexp{-2\abs{\yl}}-\xl)
\end{equation}
is the (collinear) parton luminosity,
given by the two PDFs with momentum fractions $\hat x_{1,2}= \sqrt{\frac{\tau}{x}} \myexp{\pm (Y-y)}$,
and including a $\theta$ function which is encodes the condition $\hat x_{1,2}\leq1$.

Eq.~\eqref{eq:collfactdiff} can also be rewritten as an integral over the parton momenta $\hat x_1,\hat x_2$,
which represents the direct extension of the analogous formula in DIS.
However, this form is more suitable for further manipulations.
Indeed, it has the form of a Mellin-Fourier convolution, which implies that it can be diagonalized
by taking a Mellin-Fourier transform with respect to $\tau$ and $Y$,
\begin{align}\label{eq:MellFourColl}
\int_0^1\dd\tau\, \tau^{N-1}\int_{-\infty}^\infty &\dd Y\,e^{ibY}\frac{\dd\sigma}{\dd Q^2\dd Y\dd \qt^2}
= \sum_{ij}
\frac{\dd C_{ij}}{\dd Q^2\dd y\dd\qt^2}\(N,Q^2,b,\qt^2,\as, \frac{Q^2}{\muf^2}\)\, L_{ij}\(N,b,\muf^2\),
\end{align}
where
\begin{align}\label{eq:MellFourCL}
\frac{\dd C_{ij}}{\dd Q^2\dd y\dd\qt^2}\(N,Q^2,b,\qt^2,\as, \frac{Q^2}{\muf^2}\)
  &= \int_0^1\dd x\,x^N \int_{-\infty}^\infty \dd y\,e^{iby}\frac{\dd C_{ij}}{\dd Q^2\dd y\dd\qt^2} \(x,Q^2,y,\qt^2,\as, \frac{Q^2}{\muf^2}\) \nonumber\\
L_{ij}\(N,b,\muf^2\) &= \int_0^1\dd \xl\,\xl^N \int_{-\infty}^\infty \dd \yl\,e^{ib\yl} L_{ij}\(\xl,\yl,\muf^2\) \nonumber\\
&= f_i\(N+i\frac b2,\muf^2\)\, f_j\(N-i\frac b2,\muf^2\).
\end{align}
In the last equality we have used the definition Eq.~\eqref{eq:lumi}
and changed variable from $\xl,\yl$ to $\hat x_{1,2}=\sqrt{\xl}\myexp{\pm\yl}$ and used
explicitly the $\theta$ function to obtain the product of two Mellin transforms
\beq
f_i(N,\muf^2) = \int_0^1\dd \hat x_{1,2}\,\hat x_{1,2}^N f_i(\hat x_{1,2},\muf^2).
\eeq
We further observe that the dependence on the transverse momentum does not affect the structure of the cross section formula,
and thus impacts only the kinematics.

\subsection{Extension of $\kt$ factorization to differential observables in $pp$ collisions}

The works of Refs.~\cite{Caola:2010kv,Forte:2015gve,Muselli:2017fuf} provide
a proof of a resummation formula for differential observable at LL accuracy with fixed coupling
through the so-called ladder-expansion approach.
This may seem somewhat different from the original works~\cite{Catani:1990xk,Catani:1990eg,Catani:1994sq}
where the resummation is obtained by proving a $\kt$ factorization
and comparing it with the standard collinear factorization formula.
In fact, despite the different languages, the two approaches are based exactly on the same underlying
factorization property and lead to exactly the same result.
It is thus natural to imagine that the results of Refs.~\cite{Caola:2010kv,Forte:2015gve,Muselli:2017fuf}
on differential distributions could be reformulated in terms of the $\kt$ factorization approach.

Indeed, it is not difficult to follow the steps of the derivation of
Refs.~\cite{Caola:2010kv,Forte:2015gve,Muselli:2017fuf}
and recognise the ingredients of $\kt$ factorization to construct a factorized formula.
Here, rather than repeating such a derivation, we limit ourselves to
formulate the result in $\kt$ factorization, showing that it corresponds to the results
of Refs.~\cite{Caola:2010kv,Forte:2015gve,Muselli:2017fuf} at LL and fixed coupling.

Similarly to the inclusive case, the differential cross section in $\kt$ factorization
turns out to be a straightforward extension of the collinear factorization Eq.~\eqref{eq:collfactdiff}
where the partons are replaced by off-shell gluons and integration over this offshellness is added.
The result reads
\beq\label{eq:ktfact}
\frac{\dd\sigma}{\dd Q^2\dd Y\dd \qt^2} = \tau\int_\tau^1\frac{\dd z}z\int \dd\eta \int_0^\infty \dd\xi_1 \int_0^\infty \dd\xi_2\,
\frac{\dd{\cal C}}{\dd Q^2\dd\eta\dd\qt^2}(z,\xi_1,\xi_2,Q^2,\eta,\qt^2,\as)\, \Lum\(\frac\tau z,Y-\eta,\xi_1,\xi_2\),
\eeq
where
\beq\label{eq:lumioff}
\Lum\(\zl,\el,\xi_1,\xi_2\) = {\cal F}_g\(\sqrt{\zl}e^{\el},\xi_1\)\, {\cal F}_g\(\sqrt{\zl}e^{-\el},\xi_2\)\, \theta\(e^{-2|\el|}-\zl\)
\eeq
and $\xi_{1,2}=\kvec^2_{1,2}/Q^2$ are the offshellness of the gluons normalized to the hard scale $Q^2$,
and $\kvec_{1,2}$ are the transverse components of the off-shell gluon momenta
(for more details on the kinematics, see App.~\ref{app:XS}).
In the expression above $\dd{\cal C}$ is the (differential) off-shell coefficient function,
representing the process-dependent hard scattering initiated by off-shell gluons.
More precisely, it corresponds to the last 2GI part (in the $t$ channel) of the amplitude squared of the process,
saturating the off-shell gluon indices with a suitable projector~\cite{Catani:1990xk,Catani:1990eg,Catani:1994sq}.
Everything else is collected into the two unintegrated gluon PDFs ${\cal F}_g$,
that include the standard collinear PDFs and the chain of emissions from the initial parton to the
last gluon (the ladder in the language of Refs.~\cite{Caola:2010kv,Forte:2015gve,Muselli:2017fuf}).
The integration variables $z$ and $\eta$ are the analog of $x$ and $y$ of Eq.~\eqref{eq:collfactdiff},
but referred to the center-of-mass frame of the off-shell partons.
More precisely, we consider as the parton-level center-of-mass frame in $\kt$-factorization
the one obtained if we set the off-shellness equal to zero, so that it is related to
the collider frame by a longitudinal boost.
More details are given in App.~\ref{app:XS}.

We now show that Eq.~\eqref{eq:ktfact} is equivalent to the result of Ref.~\cite{Muselli:2017fuf}.\footnote
{Notice that Ref.~\cite{Muselli:2017fuf} considers only the double differential distribution
in rapidity and transverse momentum, because it focusses on the Higgs production process,
where the invariant mass is clearly fixed to the Higgs mass.
However, the derivation there is general enough to be valid also for invariant mass distributions.}
First, we take the Mellin-Fourier transform of this expression with respect to $\tau$ and $Y$,
\begin{align}\label{eq:MellFourkt}
\int_0^1\dd\tau\, \tau^{N-1}\int_{-\infty}^\infty &\dd Y\,e^{ibY}\frac{\dd\sigma}{\dd Q^2\dd Y\dd \qt^2}\nonumber\\
&= \int_0^\infty \dd\xi_1 \int_0^\infty \dd\xi_2\,
\frac{\dd{\cal C}}{\dd Q^2\dd \eta\dd\qt^2}(N,\xi_1,\xi_2,Q^2,b,\qt^2,\as)\, \Lum\(N,b,\xi_1,\xi_2\),
\end{align}
with
\begin{align}
\frac{\dd{\cal C}}{\dd Q^2\dd \eta\dd\qt^2}(N,\xi_1,\xi_2,Q^2,b,\qt^2,\as)
  &= \int_0^1\dd z\,z^N \int_{-\infty}^\infty \dd \eta\,e^{ib\eta}\frac{\dd{\cal C}}{\dd Q^2\dd \eta\dd\qt^2} (z,\xi_1,\xi_2,Q^2,\eta,\qt^2,\as) \nonumber\\
\Lum\(N,b,\xi_1,\xi_2\) &= \int_0^1\dd \zl\,\zl^N \int_{-\infty}^\infty \dd \el\,e^{ib\el} \Lum\(\zl,\el,\xi_1,\xi_2\) \nonumber\\
&= {\cal F}_g\(N+i\frac b2,\xi_1\)\, {\cal F}_g\(N-i\frac b2,\xi_2\),
\end{align}
where we have used the definition Eq.~\eqref{eq:lumioff},
changed variable from $\zl,\el$ to $x_{1,2}=\sqrt{\zl}\myexp{\pm\el}$
(the longitudinal proton's momentum fractions carried by each off-shell gluon)
and used the $\theta$ function to obtain the product of two Mellin transforms
\beq
{\cal F}_g(N,\xi) = \int_0^1\dd x_{1,2}\,x_{1,2}^N {\cal F}_g(x_{1,2},\xi).
\eeq
At this point we follow Ref.~\cite{Catani:1990xk,Catani:1990eg,Catani:1994sq} to write
the unintegrated PDF as
\beq\label{eq:FgLLFC}
{\cal F}_g\(N,\xi\) = R(N,\as) \gamma(N,\as)\(\frac{Q^2}{\muf^2}\)^{\gamma(N,\as)}\xi^{\gamma(N,\as)-1} f_g(N,\muf^2),
\eeq
where $\gamma(N,\as)$ is the resummed (gluon) anomalous dimension at LL
and $R(N,\as)$ is a scheme dependent factor.
Note that we are ignoring quark contributions for simplicity
(we will discuss quarks later in section~\ref{sec:channels}).
Plugging Eq.~\eqref{eq:FgLLFC} into Eq.~\eqref{eq:MellFourkt}
we immediately recover the result of Ref.~\cite{Muselli:2017fuf}.
Integrating over $\qt^2$ we also reproduce the result of Ref.~\cite{Caola:2010kv}.

To reproduce the result of Ref.~\cite{Forte:2015gve}, which is not differential in rapidity,
it is simpler to integrate Eq.~\eqref{eq:ktfact} over $Y$ and then take simply a Mellin transform
before using Eq.~\eqref{eq:FgLLFC}.
The first step leads to
\beq\label{eq:ktfactnonY}
\frac{\dd\sigma}{\dd Q^2\dd \qt^2} = \tau\int_\tau^1\frac{\dd z}z \int_0^\infty \dd\xi_1 \int_0^\infty \dd\xi_2\,
\frac{\dd{\cal C}}{\dd Q^2\dd\qt^2}(z,\xi_1,\xi_2,Q^2,\qt^2,\as)\, \Lum\(\frac\tau z,\xi_1,\xi_2\),
\eeq
with
\begin{align}\label{eq:lumioffnonY}
\Lum\(\zl,\xi_1,\xi_2\)
  &= \int \dd\el\, {\cal F}_g(\sqrt{\zl}e^{\el},\xi_1)\, {\cal F}_g(\sqrt{\zl}e^{-\el},\xi_2)\, \theta(e^{-2|\el|}-\zl) \nonumber\\
  &= \int_{\zl}^1 \dd x_2\, {\cal F}_g\(\frac{\zl}{x_2},\xi_1\)\, {\cal F}_g(x_2,\xi_2).
\end{align}
Because this new rapidity-integrated luminosity has the form of a Mellin convolution,
after taking a Mellin transform of the cross section we get
\beq\label{eq:Mellkt}
\int_0^1\dd\tau\, \tau^{N-1}\frac{\dd\sigma}{\dd Q^2\dd \qt^2}
= \int_0^\infty \dd\xi_1 \int_0^\infty \dd\xi_2\,
\frac{\dd{\cal C}}{\dd Q^2\dd\qt^2}(N,\xi_1,\xi_2,Q^2,\qt^2,\as)\, {\cal F}_g\(N,\xi_1\)\, {\cal F}_g\(N,\xi_2\),
\eeq
with
\begin{align}
\frac{\dd{\cal C}}{\dd Q^2\dd\qt^2}(N,\xi_1,\xi_2,Q^2,\qt^2,\as)
  &= \int_0^1\dd z\,z^N \frac{\dd{\cal C}}{\dd Q^2\dd\qt^2} (z,\xi_1,\xi_2,Q^2,\qt^2,\as).
\end{align}
Plugging now Eq.~\eqref{eq:FgLLFC} into Eq.~\eqref{eq:Mellkt} we finally obtain the result of Ref.~\cite{Forte:2015gve}.

Because the unintegrated PDF depends on $\xi$ through $\xi^{\gamma-1}$,
the integrals over $\xi_{1,2}$ take the form of Mellin transforms.
Therefore, the results above can be expressed (up to factors)
as the $\gamma$'th Mellin moments with respect to $\xi_{1,2}$
of the partonic off-shell coefficient functions, usually called impact factors.
These results can be further supplemented with running coupling effects
as described in Refs.~\cite{Ball:2007ra,Altarelli:2008aj}.
However, as anticipated, adding running coupling effects to the impact factors
is not suitable for numerical implementation.
In the next section we will start again from Eq.~\eqref{eq:ktfact} to construct
a resummed expression at differential level in the \hell\ language,
which makes the inclusion of running coupling effects straightforward
and leads to a stable numerical implementation.

\subsection{Small-$x$ resummation of differential distributions in the \texttt{HELL} language}

The main advantage of the formulation of small-$x$ resummation of Refs.~\cite{Bonvini:2016wki,Bonvini:2017ogt,Bonvini:2018iwt}
used in the \hell\ code is the much simpler and reliable numerical implementation.
The reason is twofold.
On the one hand, the inclusion of running coupling effects in the resummation
can be done straightforwardly without approximation and without affecting the numerical performance,
as opposed to the impact-factor approach of Refs.~\cite{Ball:2007ra,Altarelli:2008aj}
where it leads to a divergent series that has to be treated in an approximate way.
On the other hand, the result can be expressed in terms of the off-shell coefficient function
directly in momentum space, as opposed to the impact-factor formulation where a double Mellin
transform in both $z$ and $\xi$ is required for each initial-state off-shell gluon.
If these Mellin transforms can be computed analytically, the (very minor) price to pay of the \hell\ formulation
is that the $\xi$ integration has to be performed numerically.
However, when the Mellin transform in $\xi$ cannot be computed analytically,
the impact-factor formulation becomes problematic, while in the \hell\ approach this does not represent a problem.

The key step of the \hell\ approach is to write the unintegrated PDF in terms
of the collinear gluon and quark-singlet PDFs in a way that includes running coupling effects.
The generic form of such an expression, valid at least at LL, is~\cite{Bonvini:2016wki,Bonvini:2017ogt,Bonvini:2018iwt}
\beq\label{eq:Foff}
{\cal F}_g(N,\xi) = U'\(N,Q^2\xi,\muf^2\) f_g(N,\muf^2) + \frac{C_F}{C_A} \[U'\(N,Q^2\xi,\muf^2\) - \delta(\xi)\] f_q(N,\muf^2),
\eeq
where
\beq
U'\(N,Q^2\xi,\muf^2\) \equiv \frac{\dd}{\dd\xi}U\(N,Q^2\xi,\muf^2\)
\eeq
and $U(N,\kvec^2,\muf^2)$ is the evolution function of the collinear gluon\footnote
{It is worth noting that the quark part of \eqref{eq:Foff} uses the same evolutor of the gluon part.
  This is justified as, in the $x\rightarrow 0$ limit, the leading splitting functions, $P_{gg}$ and $P_{gq}$,
  are identical up to a factor $\frac{C_F}{C_A}$.
  The subtraction of the $\delta\qty(\xi)$ in the quark part, that represents the no-splitting event in which the parton remains collinear,
  is required as the first splitting of the quark into a gluon must be present,
  and so that contribution must start at order $\as$.}
from the scale $\muf^2$ to the scale $\kvec^2$, times the scheme dependent function $R(N,\as)$.
In other words, $U(N,\kvec^2,\muf^2)$ is the solution of the DGLAP equation using the small-$x$ LL anomalous dimension,
which involves only gluons (they do not mix with the quarks at LL).
Keeping running coupling effects when solving the DGLAP evolution equation provides the necessary ingredient
to include the sought running coupling effects in the resummation~\cite{Bonvini:2016wki,Bonvini:2017ogt,Bonvini:2018iwt}.
Conversely, evaluating the evolution function at fixed-coupling we get back Eq.~\eqref{eq:FgLLFC}.

In practice, to simplify the numerical implementation and avoid potential numerical issues, the evolution
function is approximated in a way that reproduces exactly the results of Refs.~\cite{Ball:2007ra,Altarelli:2008aj},
namely it is valid at LL and at ``leading running coupling'' (i.e.\ leading $\beta_0$ terms are retained).
Within this approximation it takes the form~\cite{Bonvini:2017ogt,Bonvini:2018iwt}
\beq\label{eq:UABFht}
U(N,Q^2\xi,\muf^2) \simeq R(N,\as)\, D_\text{higher-twist}\(\frac{Q^2}{\muf^2}\xi\)\,  U_{\rm ABF}\(N,\frac{Q^2}{\muf^2}\xi\),
\eeq
where
\beq
D_\text{higher-twist}(\xi) =
\begin{cases}
1 & \xi\geq1\\
1-\(-\as\beta_0\log\xi\)^{1+\frac1{\as\beta_0}}\qquad & \xi_0<\xi<1 \\
0 & \xi\leq\xi_0, \qquad \xi_0=\exp\frac{-1}{\as\beta_0},
\end{cases}
\eeq
is a damping function at small $\xi$, designed to keep unaffected the perturbative expansion of the evolution function while
ensuring that it vanishes at the Landau pole $\xi_0$ as it would do at LL with full running coupling~\cite{Bonvini:2017ogt}, and
\beq\label{eq:UABF}
U_{\rm ABF}(N,\xi) = \Big(1+r(N,\as)\log\xi\Big)^\frac{\gamma(N,\as)}{r(N,\as)},\qquad\qquad
r(N,\as) = \as^2\beta_0\frac{\dd}{\dd\as}\log\[\gamma(N,\as)\],
\eeq
is the approximated evolution function.
The anomalous dimension $\gamma$ appearing above is in principle the LL anomalous dimension.
However, it is convenient to include subleading contributions in it that simply produce subleading effects in the resummation
but make the result consistent with the resummation in DGLAP evolution.
As this discussion is not central for the present work, we refer the Reader to Ref.~\cite{Bonvini:2018iwt}
for further detail.
In the numerical implementation, we will ignore the scheme factor $R(N,\as)$. The reason is that
we use to perform small-$x$ resummation in the so-called $Q_0\MSbar$ scheme~\cite{Catani:1993ww,Catani:1994sq,Ciafaloni:2005cg,Marzani:2007gk}
where by definition $R(N,\as)=1$. This scheme differs from the usual $\MSbar$ scheme
at relative order $\as^3$ (at LL), and therefore it can be safely used in conjunction with
$\MSbar$ fixed-order computations up to NNLO.

Let us focus for simplicity on the gluon contribution only, thus neglecting the quark term in Eq.~\eqref{eq:Foff}.
Plugging Eq.~\eqref{eq:Foff} into Eq.~\eqref{eq:MellFourkt} we get
\begin{align}\label{eq:MellFourkt2}
\int_0^1\dd\tau\, \tau^{N-1}\int_{-\infty}^\infty \dd Y\,e^{ibY}\frac{\dd\sigma}{\dd Q^2\dd Y\dd \qt^2}
&= \int_0^\infty \dd\xi_1 \int_0^\infty \dd\xi_2\, \frac{\dd{\cal C}}{\dd Q^2\dd \eta\dd\qt^2}(N,\xi_1,\xi_2,Q^2,b,\qt^2,\as) \nonumber\\
  &\times U'\(N+i\frac b2,Q^2\xi_1,\muf^2\)\, f_g\(N+i\frac b2,\muf^2\) \nonumber\\
  &\times U'\(N-i\frac b2,Q^2\xi_2,\muf^2\)\, f_g\(N-i\frac b2,\muf^2\).
\end{align}
Comparing this expression with the gluon-gluon channel of the collinear factorization expression Eq.~\eqref{eq:MellFourColl} and
Eq.~\eqref{eq:MellFourCL} we find the identification
\begin{align}\label{eq:resCggNb}
  \frac{\dd C_{gg}}{\dd Q^2\dd y\dd\qt^2}\(N,Q^2,b,\qt^2,\as, \frac{Q^2}{\muf^2}\)
&= \int_0^\infty \dd\xi_1 \int_0^\infty \dd\xi_2\, \frac{\dd{\cal C}}{\dd Q^2\dd \eta\dd\qt^2}(N,\xi_1,\xi_2,Q^2,b,\qt^2,\as) \nonumber\\
  &\times U'\(N+i\frac b2,Q^2\xi_1,\muf^2\) U'\(N-i\frac b2,Q^2\xi_2,\muf^2\).
\end{align}
So far this is not dissimilar to the approach of older works; in particular, if one replaces $U'$
with the LL fixed-coupling expression from Eq.~\eqref{eq:FgLLFC} one recognises the
definition of the impact fator.
Here instead, we keep a more generic expression for $U'$ and further manipulate the result.
Indeed, we notice that the $N,b$ dependence of the right-hand side of Eq.~\eqref{eq:resCggNb}
has the same form of the right-hand side of Eq.~\eqref{eq:MellFourColl} or Eq.~\eqref{eq:MellFourkt}.
We thus recognise Eq.~\eqref{eq:resCggNb} as the Mellin-Fourier transform of
\begin{align}\label{eq:resCggzy}
  \frac{\dd C_{gg}}{\dd Q^2\dd y\dd\qt^2}\(x,Q^2,y,\qt^2,\as, \frac{Q^2}{\muf^2}\)
  &= \int_0^\infty \dd\xi_1 \int_0^\infty \dd\xi_2 \int_x^1\frac{\dd z}z \int_{-\frac12\log\frac zx}^{\frac12\log\frac zx} \dd\el\nonumber\\
  &\times \frac{\dd{\cal C}}{\dd Q^2\dd \eta\dd\qt^2}(z,\xi_1,\xi_2,Q^2,y-\el,\qt^2,\as) \nonumber\\
  &\times U'\(\sqrt{\frac xz}e^{\el},Q^2\xi_1,\muf^2\) U'\(\sqrt{\frac xz}e^{-\el},Q^2\xi_2,\muf^2\),
\end{align}
which is expressed as a 4-dimensional integral (to be performed numerically in general) over simple quantities,
namely the differential off-shell coefficient function 
and the evolution factors in physical momentum space.
This result is very convenient from a numerical point of view.
The two additional integrations over $z$ and $\eta$ are much simpler to compute
than the inverse Mellin-Fourier transform over $N$ and $b$ of Eq.~\eqref{eq:resCggNb},
especially in \hell, because the anomalous dimension appearing in the definition of $U'$
is available in \hell\ only for values of $N$ along a specific inversion contour, which would not be sufficient
here due to the $\pm ib$ imaginary shift.
Instead, because the evolution function $U$ is universal (process independent),
it is computed once and for all in \hell\ directly in momentum space,
and it can be easily used in an expression like Eq.~\eqref{eq:resCggzy}.
Moreover, as already mentioned, with respect to the impact-factor formulation
this result easily incorporates the running coupling contributions through the
use of the proper evolution function $U$, Eq.~\eqref{eq:UABFht}.

We want to emphasize a difference with respect to previous formulations of resummation in the \hell\ language.
In previous works, because the $N$ dependence of the off-shell coefficient function is subleading,
we used to set $N=0$ in it before computing the inverse Mellin transform.
The main motivation was that the analytical expressions obtained in this way were simpler,
and in some cases it is not possible to compute the Mellin transform of the off-shell coefficient function
analytically for generic $N$, but it is possible for $N=0$.
In our case, this approach would correspond to setting $N=0$ in the off-shell coefficient function in Eq.~\eqref{eq:resCggNb}
before computing the inverse Mellin-Fourier transform.
However, when dealing with differential distribution we are often not able to compute analytically the Mellin transform of the off-shell
coefficient function, not even in $N=0$. So there would be no advantage in setting $N=0$ in it.
Conversely, there would be disadvantages. Indeed, some physical kinematic constraints
would be approximated if computed in $N=0$. One of the consequences is that the endpoint
of the rapidity distribution, which is a physical property of the process determined by its kinematics,
would be wrong when setting $N=0$.
This is not dissimilar to what has been found in Ref.~\cite{Bonvini:2017ogt}
in the case of DIS, where quark mass effects on kinematic constraints were lost
when setting $N=0$, requiring a restoration of the constraints by hand.
Here, we thus decide that it is much better (and simpler) to keep the subleading $N$ dependence,
thereby preserving physical kinematic constraints, without paying any price from the numerical point of view.

\subsection{All partonic channels}
\label{sec:channels}

In the resummed expression Eq.~\eqref{eq:resCggzy} the key ingredient is (the $\xi$-derivative of)
the evolution function in $x$ space,\footnote
{As we are running out of letters, we are now using $x$ for the generic first argument of the evolution function in momentum space,
  not to be confused with the variable $x=Q^2/\hat s_c$ which is the argument of the collinear coefficient function.}
computed in \hell\ as the inverse Mellin transform of Eq.~\eqref{eq:UABFht}.
We observe that such inverse Mellin transform is a distribution.
Indeed, expanding $U(N,Q^2\xi,\muf^2)$ in powers of $\as$ the zeroth order term is just $1$, whose inverse Mellin is $\delta(1-x)$.
Since this is the only distributional contribution in $U$, we find it more convenient to write it explicitly,
\beq\label{eq:Ureg}
U(N, Q^2\xi,\muf^2) = 1+U_{\rm reg}(N, Q^2\xi,\muf^2) \qquad\Leftrightarrow\qquad 
U(x, Q^2\xi,\muf^2) = \delta(1-x)+U_{\rm reg}(x, Q^2\xi,\muf^2),
\eeq
where $U_{\rm reg}$ is an ordinary function.
Computing the $\xi$-derivative appearing in Eq.~\eqref{eq:Foff} is not entirely trivial.
To do so we first introduce explicitly a factor $\theta(\xi)$ in the definition of the evolution function,
$U(N, Q^2\xi,\muf^2) = \theta(\xi)\[1+U_{\rm reg}(N, Q^2\xi,\muf^2)\]$,
which is conceptually harmless as certainly the scale $Q^2\xi=\kvec^2$ has to be positive.
When deriving we get
\begin{align}\label{eq:U'regN}
  U'(N, Q^2\xi,\muf^2)
  &=\delta(\xi) + \delta(\xi) U_{\rm reg}(N, 0,\muf^2) + \theta(\xi)U'_{\rm reg}(N, Q^2\xi,\muf^2) \nonumber\\
  &=\delta(\xi) - \delta(\xi) \int_0^{\frac{\muf^2}{Q^2}} \dd\xi'\,U'_{\rm reg}(N, Q^2\xi',\muf^2) + \theta(\xi)U'_{\rm reg}(N, Q^2\xi,\muf^2) \nonumber\\
  &=\delta(\xi) + \[U'_{\rm reg}(N, Q^2\xi,\muf^2)\]_+
\end{align}
where in the second step we have used the fact that $U_{\rm reg}(N, \muf^2,\muf^2)=0$
and in the last step we have defined the plus distribution according to
\beq
\int_0^\infty \dd\xi\,f(\xi)\[g(\xi)\]_+ = 
\int_0^{\frac{\muf^2}{Q^2}} \dd\xi\,\[f(\xi)-f(0)\]g(\xi) +
\int_{\frac{\muf^2}{Q^2}}^\infty \dd\xi\,f(\xi)g(\xi) .
\eeq
The $\delta(\xi)$ term appearing as the derivative of the zeroth order of the evolution
has a precise physical meaning: it represents the undisturbed gluon, that does not emit
and thus it remains on-shell ($\xi=0$). This indeed corresponds to the term subtracted
in the quark contribution to the unintegrated PDF, Eq.~\eqref{eq:Foff}.

We now observe that the introduction of the plus distribution is not really necessary,
because the contribution $U_{\rm reg}(N, 0,\muf^2)$ appearing in the first line of Eq.~\eqref{eq:U'regN}
is finite. More precisely, because $U(N, 0,\muf^2)=0$ by construction, Eq.~\eqref{eq:UABFht}, we have $U_{\rm reg}(N, 0,\muf^2) =-1$,
corresponding in $x$-space to
\beq\label{eq:Uregxi0}
U_{\rm reg}(x, 0,\muf^2) = -\delta(1-x).
\eeq
If this is the case, the first two terms in the first line of Eq.~\eqref{eq:U'regN} would cancel,
thus leaving the simpler result $U'(N, Q^2\xi,\muf^2)=U'_{\rm reg}(N, Q^2\xi,\muf^2)$ which is what we would have obtained
if we hadn't introduced the $\theta$ function.
This implies that the nice physical distinction between the no-emission contribution $\delta(\xi)$ and the
at-least-one-emission contribution $U'_{\rm reg}(N, Q^2\xi,\muf^2)$ gets lost.
This is clearly undesirable, and may hint at a problem in the construction of the evolution function.

To understand and overcome this problem, we observe that the $\xi\to0$ limit of $U'_{\rm reg}$, Eq.~\eqref{eq:Uregxi0},
is localised at large $x$. But the evolution function at large $x$ is not
expected to be accurate, as it is constructed to resum logarithmic contributions
at small $x$. Therefore, we can (and we do) damp the function $U_{\rm reg}(x,Q^2\xi,\muf^2)$
(and thus its $\xi$-derivative) at large $x$, with a damping function of the form $(1-x)^a$
(we use $a=2$ in the code). After damping, the evolution function satisfies
\beq\label{eq:Uregx1}
U_{\rm reg}(x=1, Q^2\xi,\muf^2) = 0,
\eeq
for any value of $\xi$, including $\xi=0$.
In this way, we obtain $U_{\rm reg}(x, 0,\muf^2) =0$ and thus $U_{\rm reg}(N, 0,\muf^2) =0$,
implying that the second term in the first line of Eq.~\eqref{eq:U'regN} vanishes, thus giving
\beq\label{eq:U'reg}
U'(N, Q^2\xi,\muf^2) = \delta(\xi)+U'_{\rm reg}(N, Q^2\xi,\muf^2) \qquad\Leftrightarrow\qquad 
U'(x, Q^2\xi,\muf^2) = \delta(\xi)\delta(1-x)+U'_{\rm reg}(x, Q^2\xi,\muf^2).
\eeq
In other words, because of the large-$x$ damping, the plus distribution is ineffective.
For completeness, we have verified that the numerical integral of $U'_{\rm reg}(x, Q^2\xi,\muf^2)$
from zero to $\muf^2/Q^2$ gives indeed zero for all values of $x$.

Let us now come back to the resummed coefficient function.
According to Eq.~\eqref{eq:U'reg}, the unintegrated PDF Eq.~\eqref{eq:Foff} can be rewritten as
\beq\label{eq:Foff2}
{\cal F}_g(N,\xi) = \[U'_{\rm reg}\(N,Q^2\xi,\muf^2\)+\delta(\xi)\] f_g(N,\muf^2) + \frac{C_F}{C_A} U'_{\rm reg}\(N,Q^2\xi,\muf^2\) f_q(N,\muf^2).
\eeq
Physically, the $\delta(\xi)$ contribution in the equation above represents the (on-shell) gluon that does not emit,
thus producing no logs: this is the fixed-order contribution, and it reproduces the on-shell result.
The other term, $U'_{\rm reg}$, is the term containing at least one emission, and thus at least one small-$x$ log.

Starting from Eq.~\eqref{eq:Foff2} and proceeding as in the previous section, keeping also the quark contributions this time,
we obtain the following expressions
\begin{subequations}\label{eq:resC}
\begin{align}
  \frac{\dd C_{gg}}{\dd Q^2\dd y\dd\qt^2}\(x,Q^2,y,\qt^2,\as, \frac{Q^2}{\muf^2}\)
  &= \int_0^\infty \dd\xi_1 \int_0^\infty \dd\xi_2 \int_x^1\frac{\dd z}z \int_{-\frac12\log\frac zx}^{\frac12\log\frac zx} \dd\bar\eta\,\nonumber\\
  &\times \frac{\dd{\cal C}}{\dd Q^2\dd \eta\dd\qt^2}(z,\xi_1,\xi_2,Q^2,y-\bar\eta,\qt^2,\as) \nonumber\\
  &\times \[U'_{\rm reg}\(\sqrt{\frac xz}e^{\bar\eta},Q^2\xi_1,\muf^2\)+\delta(\xi_1)\delta\(1-\sqrt{\frac xz}e^{\bar\eta}\)\]\nonumber\\
  &\times \[U'_{\rm reg}\(\sqrt{\frac xz}e^{-\bar\eta},Q^2\xi_2,\muf^2\) +\delta(\xi_2)\delta\(1-\sqrt{\frac xz}e^{-\bar\eta}\)\], \\
  \frac{\dd C_{qg}}{\dd Q^2\dd y\dd\qt^2}\(x,Q^2,y,\qt^2,\as, \frac{Q^2}{\muf^2}\)
  &= \frac{C_F}{C_A}\int_0^\infty \dd\xi_1 \int_0^\infty \dd\xi_2 \int_x^1\frac{\dd z}z \int_{-\frac12\log\frac zx}^{\frac12\log\frac zx} \dd\bar\eta\,\nonumber\\
  &\times \frac{\dd{\cal C}}{\dd Q^2\dd \eta\dd\qt^2}(z,\xi_1,\xi_2,Q^2,y-\bar\eta,\qt^2,\as) \nonumber\\
  &\times U'_{\rm reg}\(\sqrt{\frac xz}e^{\bar\eta},Q^2\xi_1,\muf^2\)\nonumber\\
  &\times \[U'_{\rm reg}\(\sqrt{\frac xz}e^{-\bar\eta},Q^2\xi_2,\muf^2\) +\delta(\xi_2)\delta\(1-\sqrt{\frac xz}e^{-\bar\eta}\)\], \\
  \frac{\dd C_{gq}}{\dd Q^2\dd y\dd\qt^2}\(x,Q^2,y,\qt^2,\as, \frac{Q^2}{\muf^2}\)
  &= \frac{C_F}{C_A}\int_0^\infty \dd\xi_1 \int_0^\infty \dd\xi_2 \int_x^1\frac{\dd z}z \int_{-\frac12\log\frac zx}^{\frac12\log\frac zx} \dd\bar\eta\,\nonumber\\
  &\times \frac{\dd{\cal C}}{\dd Q^2\dd \eta\dd\qt^2}(z,\xi_1,\xi_2,Q^2,y-\bar\eta,\qt^2,\as) \nonumber\\
  &\times \[U'_{\rm reg}\(\sqrt{\frac xz}e^{\bar\eta},Q^2\xi_1,\muf^2\)+\delta(\xi_1)\delta\(1-\sqrt{\frac xz}e^{\bar\eta}\)\]\nonumber\\
  &\times U'_{\rm reg}\(\sqrt{\frac xz}e^{-\bar\eta},Q^2\xi_2,\muf^2\), \\
  \frac{\dd C_{qq}}{\dd Q^2\dd y\dd\qt^2}\(x,Q^2,y,\qt^2,\as, \frac{Q^2}{\muf^2}\)
  &= \(\frac{C_F}{C_A}\)^2\int_0^\infty \dd\xi_1 \int_0^\infty \dd\xi_2 \int_x^1\frac{\dd z}z \int_{-\frac12\log\frac zx}^{\frac12\log\frac zx} \dd\bar\eta\,\nonumber\\
  &\times \frac{\dd{\cal C}}{\dd Q^2\dd \eta\dd\qt^2}(z,\xi_1,\xi_2,Q^2,y-\bar\eta,\qt^2,\as) \nonumber\\
  &\times U'_{\rm reg}\(\sqrt{\frac xz}e^{\bar\eta},Q^2\xi_1,\muf^2\)\, U'_{\rm reg}\(\sqrt{\frac xz}e^{-\bar\eta},Q^2\xi_2,\muf^2\).
\end{align}
\end{subequations}
These results can be written in a more compact form as
\begin{subequations}\label{eq:resC_reg_aux}
\begin{align}
\frac{\dd C_{gg}}{\dd Q^2\dd y\dd\qt^2}
&= \frac{\dd C_{\rm reg}}{\dd Q^2\dd y\dd\qt^2} + \frac{\dd C_{\rm aux\,+}}{\dd Q^2\dd y\dd\qt^2} + \frac{\dd C_{\rm aux\,-}}{\dd Q^2\dd y\dd\qt^2} 
  +\frac{\dd {\cal C}}{\dd Q^2\dd \eta\dd\qt^2}(x,0,0,Q^2,y,\qt^2,\as), \\
\frac{\dd C_{qg}}{\dd Q^2\dd y\dd\qt^2}
&= \frac{C_F}{C_A} \[ \frac{\dd C_{\rm reg}}{\dd Q^2\dd y\dd\qt^2} + \frac{\dd C_{\rm aux\,+}}{\dd Q^2\dd y\dd\qt^2}\], \\
\frac{\dd C_{gq}}{\dd Q^2\dd y\dd\qt^2}
&= \frac{C_F}{C_A} \[ \frac{\dd C_{\rm reg}}{\dd Q^2\dd y\dd\qt^2} + \frac{\dd C_{\rm aux\,-}}{\dd Q^2\dd y\dd\qt^2}\], \\
\frac{\dd C_{qq}}{\dd Q^2\dd y\dd\qt^2}
&= \(\frac{C_F}{C_A}\)^2 \frac{\dd C_{\rm reg}}{\dd Q^2\dd y\dd\qt^2}
\end{align}
\end{subequations}
having defined
\begin{align}\label{eq:resCreg}
  \frac{\dd C_{\rm reg}}{\dd Q^2\dd y\dd\qt^2}\(x,Q^2,y,\qt^2,\as, \frac{Q^2}{\muf^2}\)
  &= \int_0^\infty \dd\xi_1 \int_0^\infty \dd\xi_2 \int_x^1\frac{\dd z}z \int_{-\frac12\log\frac zx}^{\frac12\log\frac zx} \dd\bar\eta\,\nonumber\\
  &\times \frac{\dd{\cal C}}{\dd Q^2\dd \eta\dd\qt^2}(z,\xi_1,\xi_2,Q^2,y-\bar\eta,\qt^2,\as) \nonumber\\
  &\times U'_{\rm reg}\(\sqrt{\frac xz}e^{\bar\eta},Q^2\xi_1,\muf^2\)\, U'_{\rm reg}\(\sqrt{\frac xz}e^{-\bar\eta},Q^2\xi_2,\muf^2\)
\end{align}
and
\begin{align}\label{eq:resCaux}
  \frac{\dd C_{\rm aux\,\pm}}{\dd Q^2\dd y\dd\qt^2}\(x,Q^2,y,\qt^2,\as, \frac{Q^2}{\muf^2}\)
  &= \int_0^\infty \dd\xi \int_x^1\frac{\dd z}z 
  \frac{\dd{\cal C}}{\dd Q^2\dd \eta\dd\qt^2}\(z,\xi,0,Q^2,y\pm\frac12\log\frac zx,\qt^2,\as\) \nonumber\\
  &\times U'_{\rm reg}\(\frac xz,Q^2\xi,\muf^2\),
\end{align}
where in the last equation we have used the symmetry $\xi_1\leftrightarrow\xi_2$ of the off-shell coefficient.
So in conclusion the resummed expressions for all channels are written in terms of a ``regular'' resummed coefficient
and two simpler ``auxiliary'' functions,\footnote
{The name ``auxiliary'' follows the nomenclature introduced in Ref.~\cite{Bonvini:2018iwt},
  extended to differential distributions.}
each defined in terms of integrals over ordinary functions (and thus easy to implement numerically).
The $gg$ coefficient function also depends on the on-shell limit of the off-shell coefficient;
however, whenever the resummed result is matched to a fixed-order computation, this contribution will be subtracted
and thus in practical applications it will never be needed.

We observe that the auxiliary functions are obtained by putting on shell one of the incoming gluons.
Therefore, they represent a contribution in which resummation, obtained from $\kt$ factorization,
acts on a single initial state parton, while the other obeys the standard collinear factorization.
This resembles the hybrid factorization discussed in Refs.~\cite
{Deak:2009xt, Deak:2011ga, Celiberto:2020tmb, Celiberto:2022rfj, Celiberto:2022dyf, vanHameren:2022mtk}
and used to describe forward production.
We believe that our auxiliary contribution does indeed represent the same
resummed contributions obtained from the hybrid factorization. However, there may be
some differences due to the different approaches to resummation,
that we aim at investigating in a future work.

\subsection{Matching to fixed order}
\label{sec:matching}

The resummed result Eq.~\eqref{eq:resC} contains only the small-$x$ logarithms.
For phenomenological applications, it has to be matched with a fixed-order computation.
To do this, we need to compute its expansion in powers of $\as$ up to some order, subtract it and replace it
with the exact fixed-order result at the same order.

Computing the $\as$ expansion of the resummed result is in principle straightforward,
but it needs some care in practice as we shall now see.
Note that the $\as$ dependence comes fully from the integrand of Eq.~\eqref{eq:resC}, and specifically from the evolution function
$U'_{\rm reg}$, as the off-shell coefficient function is needed only at the lowest non-trivial order to achieve LL accuracy.
To construct the expansion of $U'_{\rm reg}$, let us consider the expansion of $U_{\rm reg}$ first.
Because of Eqs.~\eqref{eq:UABFht} and \eqref{eq:UABF}, it is clear that such an expansion contains powers of $\log\xi$.
The first couple of orders take the form (in both the $\MSbar$ and $Q_0\MSbar$ schemes)
\begin{align}\label{eq:Uregexp}
U_{\rm reg}(N,Q^2\xi,\muf^2) &= \as(\muf^2)\gamma_0(N)\log\frac{Q^2\xi}{\muf^2} \nonumber\\
&+ \as^2(\muf^2)\[\gamma_1(N)\log\frac{Q^2\xi}{\muf^2} + \frac12\gamma_0(N)\(\gamma_0(N)-\beta_0\)\log^2\frac{Q^2\xi}{\muf^2}\] + \Ord(\as^3),
\end{align}
having assumed the expansion $\gamma(N,\as) = \as\gamma_0+\as^2\gamma_1+\Ord(\as^3)$ for the resummed anomalous dimension
(see Refs.~\cite{Bonvini:2018iwt,Bonvini:2018xvt} for explicit expressions).
After computing the derivative with respect to $\xi$, terms of the form $\log^k\xi/\xi$ appear.
Such terms are not integrable in the $\xi\to0$ limit, and thus require a regularization procedure.

To do so, we recall that the actual form of the derivative of the evolution function
has a plus distribution around $U'_{\rm reg}$, Eq.~\eqref{eq:U'regN}.
The plus distribution does not play a role at resummed level
because to all orders $U'_{\rm reg}(N,0,\muf^2)=0$, but this is not true order by order.
The order by order expansion of the evolution function diverges at $\xi=0$,
and so the plus distribution becomes essential.

With a slight abuse of notation,\footnote
{The most correct way of writing these results is to keep the plus distribution
around $U'_{\rm reg}$ everywhere.}
starting from Eq.~\eqref{eq:Uregexp},
we can write the first couple of orders of the expansion of $U'_{\rm reg}$,
\begin{align}\label{eq:U'regexp}
U'_{\rm reg}(N,Q^2\xi,\muf^2) &= \as(\muf^2)\gamma_0(N)\(\frac1{\xi}\)_+ \nonumber\\
&+ \as^2(\muf^2)\[\gamma_1(N) \(\frac1{\xi}\)_+ + \gamma_0(N)\(\gamma_0(N)-\beta_0\) \(\frac{\log\frac{Q^2\xi}{\muf^2}}{\xi}\)_+\] + \Ord(\as^3),
\end{align}
or, in $x$ space,
\begin{align}\label{eq:U'regexpx}
U'_{\rm reg}(x,Q^2\xi,\muf^2) &= \as(\muf^2)P_0(x)\(\frac1{\xi}\)_+ \nonumber\\
&+ \as^2(\muf^2)\[P_1(x) \(\frac1{\xi}\)_+ + \(P_{00}(x)-\beta_0P_0(x)\) \(\frac{\log\frac{Q^2\xi}{\muf^2}}{\xi}\)_+\] + \Ord(\as^3),
\end{align}
having defined $P_{00}(x)$ as the Mellin convolution of two $P_0$'s,
and having used the expansion $P(x,\as) = \as P_0(x) +\as^2P_1(x)+\Ord(\as^3)$
which is the inverse Mellin transform of the resummed anomalous dimension $\gamma(N,\as)$.

Plugging these expansions into Eq.~\eqref{eq:resCreg} and Eq.~\eqref{eq:resCaux}
we finally obtain the sought perturbative expansion of the resummed result.
In particular, we find up to relative $\Ord(\as^2)$
\begin{align}
  \frac{\dd C_{\rm reg}}{\dd Q^2\dd y\dd\qt^2}\(x,Q^2,y,\qt^2,\as, \frac{Q^2}{\muf^2}\)
  &= \int_0^\infty \dd\xi_1 \int_0^\infty \dd\xi_2 \int_x^1\frac{\dd z}z \int_{-\frac12\log\frac zx}^{\frac12\log\frac zx} \dd\bar\eta\,\nonumber\\
  &\times \frac{\dd{\cal C}}{\dd Q^2\dd \eta\dd\qt^2}(z,\xi_1,\xi_2,Q^2,y-\bar\eta,\qt^2,\as) \nonumber\\
  &\times \[\as^2(\muf^2) \(\frac1{\xi_1}\)_+ \(\frac1{\xi_2}\)_+ P_0\(\sqrt{\frac xz}e^{\bar\eta}\)\, P_0\(\sqrt{\frac xz}e^{-\bar\eta}\) +\Ord(\as^3)\]\label{eq:resCregFO} \\
  \frac{\dd C_{\rm aux\,\pm}}{\dd Q^2\dd y\dd\qt^2}\(x,Q^2,y,\qt^2,\as, \frac{Q^2}{\muf^2}\)
  &= \int_0^\infty \dd\xi \int_x^1\frac{\dd z}z 
  \frac{\dd{\cal C}}{\dd Q^2\dd \eta\dd\qt^2}\(z,\xi,0,Q^2,y\pm\frac12\log\frac zx,\qt^2,\as\) \nonumber\\
  &\times \bigg\{\as(\muf^2)P_0\(\frac xz\)\(\frac1{\xi}\)_+ \nonumber\\
  &\qquad+ \as^2(\muf^2)\[P_1\(\frac xz\) \(\frac1{\xi}\)_+ + \(P_{00}\(\frac xz\)-\beta_0P_0\(\frac xz\)\) \(\frac{\log\frac{Q^2\xi}{\muf^2}}{\xi}\)_+\]\nonumber\\
  &\qquad+ \Ord(\as^3)\bigg\}\label{eq:resCauxFO}
\end{align}
out of which we can construct the expansion of each coefficient function through Eq.~\eqref{eq:resC_reg_aux}.

We note in conclusion that this procedure is not dissimilar to what was used in previous works, see e.g.\ Ref.~\cite{Bonvini:2018iwt},
where the expansion was obtained by expanding the impact factor.
However, the derivation obtained here is more ``direct'',
and the result is written in a form that is immediately
usable to compute the expansion numerically, without the need to compute analytically the impact factor.

\section{Heavy-quark pair production}
\label{sec:HQ}

Having described the general formalism for the small-$x$ resummation of differential distributions
in \hell, we now focus on a specific process: heavy-quark pair production in proton-proton collisions.
This process is relevant because at the LHC, and in particular at LHCb, it is measured
in the forward region where one parton is at small $x$, and it can thus provide important
constraints on the PDFs (the gluon in particular) in a region of $x$ that is so far unexplored.
Moreover, NLO results for this process are available~\cite{Nason:1987xz,Frixione:2007nw},
and NNLO corrections have also been computed recently~\cite{Catani:2020kkl},
making this process suitable for precision studies.

The process can be schematized as
\begin{equation} \label{hfhadroproduction}
\mathrm{p}\qty(P_1) + \mathrm{p}\qty(P_2) \rightarrow Q \qty(p) + \bar Q \qty(\bar p) + X ,
\end{equation}
where the two incoming protons have light-cone momenta $P_{1,2}$ with $(P_1+P_2)^2=s$,
the outgoing heavy quarks have mass $m$ and momenta $p,\bar p$ with $p^2=\bar p^2=m^2$,
and $X$ represents any additional radiation together with the remnants of the protons.
For simplicity, we consider the final state to be given by the heavy quarks themselves,
thus ignoring their hadronization and eventual decay into lighter hadrons.\footnote
{This simplification does not raise concerns about infrared safety,
  as the mass of the heavy quarks acts as an infrared regulator for the final state.}
These effects should not affect the impact of resummation,
as they factorize (at least at LL) with respect to the hard scattering process.
A full phenomenological study of the process including these effects is beyond the scope of this paper and is left to future work.
Rather, the scope of this section is to demonstrate the application of the framework introduced in this paper.

The resummation of high-energy logarithms in heavy quark pair production has been considered in the literature,
both at the level of the total cross section~\cite{Catani:1990eg,Ball:2001pq}
and for some differential observables~\cite{Baranov:2002cf,Kniehl:2006sk,Bolognino:2019yls,Celiberto:2022rfj,Celiberto:2022dyf}.
To perform small-$x$ resummation of differential distributions in our approach,
we need to compute the coefficient function of the partonic subprocess where
two off-shell gluons produce the final state. At lowest order, as appropriate for LL resummation, the process is
\beq
g^*(k_1) + g^*(k_2) \to Q \qty(p) + \bar Q \qty(\bar p),
\eeq
where the off-shell gluon momenta are parametrized as\footnote
{Here we are using a slightly inconsistent notation.
  Indeed, we assume that the bold vectors $\kvec_{1,2}$ are 2-dimensional Euclidean vectors in the transverse plane.
  However, when they are summed to 4-dimensional Minkowski vectors, we mean them to be the 4-vector with the same spatial components.
  The confusion may only arise when they appear in a scalar product, because the two interpretations would differ by a sign.
  In these cases, we always consider them as 2-vectors.}
\begin{subequations}
\begin{align}
k_1 &= x_1 P_1 + \kvec_1, \\
k_2 &= x_2 P_2 + \kvec_2.
\end{align}
\end{subequations}
In this way, the off-shellness of the gluons is given by a transverse component with respect to the beam axis.
The longitudinal momentum fractions $x_{1,2}$ correspond to the first argument of the unintegrated PDFs,
and their ratio is related to the longitudinal boost of the partonic reference frame used to compute the coefficient function
by $Y-\eta=\frac12\log\frac{x_1}{x_2}$. Note that this frame is not in general the
partonic center-of-mass frame due to the presence of a transverse component in the gluon momenta,
but it reduces to it in the limit where the gluons are on shell.
Additional information on the kinematics is given in Appendix~\ref{app:XS}.

In order to compute the actual off-shell coefficient function, we need to decide which is the vector $q$ with respect to which we want to be differential.
There are two natural choices: either $q$ is one of the two heavy quark momenta $p$ or $\bar p$,
or it is the sum of the two momenta, thus representing the momentum of the pair.
We now present results for either choice in turn.

\subsection{Results differential in the single heavy-quark}
\label{sec:singlekin}

In this section we consider the final state to be one of the heavy quarks,
and thus focus on the differential distribution in the components of the momentum $q=p$.
The details of the computation of the partonic off-shell coefficient function are given in App.~\ref{app:singlekin}.

\begin{figure}[t]
  \centering
  \includegraphics[width=0.49\textwidth,page=9 ]{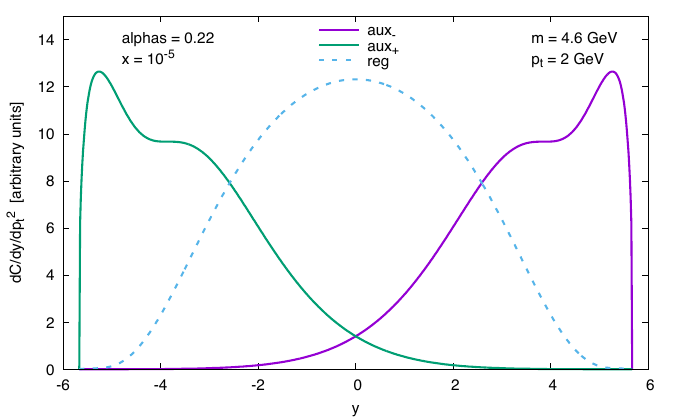}
  \includegraphics[width=0.49\textwidth,page=10]{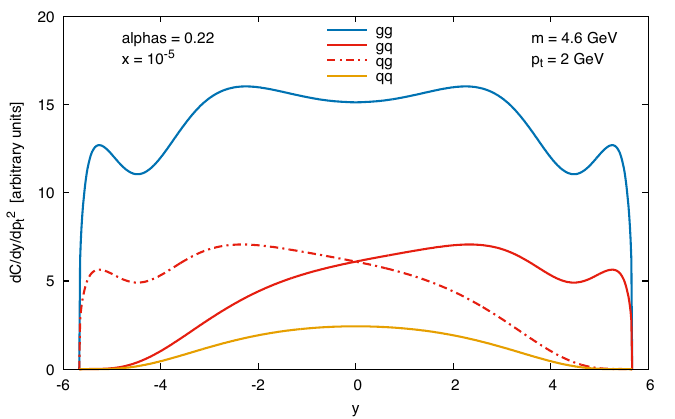}
  \caption{The auxiliary Eq.~\eqref{eq:resCaux} and regular Eq.~\eqref{eq:resCreg} functions as a function of partonic rapidity $y$
    for single quark production of mass $m=4.6$~GeV at $p_t=2$~GeV and $x=10^{-5}$ (left plot).
    The resummed coefficient functions at parton level for each partonic channel
    constructed according to Eq.~\eqref{eq:resC_reg_aux} for the same kinematics (right plot).}
  \label{fig:SQDoubleDiffPartonic}
\end{figure}

We start by presenting the resummed result at parton level, computed according to Eqs.~\eqref{eq:resC_reg_aux}.
We consider the resummed coefficient functions for bottom pair production, with $m_b=4.6$~GeV,
double differential in (partonic) rapidity $y$ and transverse momentum $p_t$ of the bottom quark.\footnote
{As we consider the bottom quarks to be on shell, the invariant mass distribution is a delta function
  and therefore for this process the triple differential distribution is of no interest.}
In Fig.~\ref{fig:SQDoubleDiffPartonic}, we show such distribution as a function of $y$ and for fixed $p_t=2$~GeV,
which is a value accessible at LHCb for the production of $b$-hadrons.
In the left panel, we plot separately the regular Eq.~\eqref{eq:resCreg} and auxiliary Eq.~\eqref{eq:resCaux} contributions
out of which the various channels can be built according to Eqs.~\eqref{eq:resC_reg_aux},
while, in the right panel, we combine them according to those equations to
construct the coefficient functions for the $gg$, $gq$, $qg$ and $qq$ channels.
We observe that the shapes of these functions are quite peculiar,
mostly due to the peak of the auxiliary contribution at large rapidity.
However, we stress that these are parton-level results, and they
are not expected to behave smoothly. In fact, due to the all-order nature
of these contributions, it is natural that they present some new features missing in the fixed order.

\begin{figure}[t]
  \centering
  \includegraphics[width=0.49\textwidth,page=3]{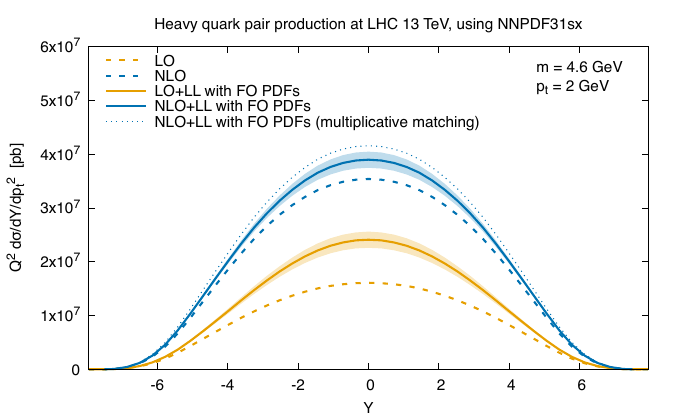}
  \includegraphics[width=0.49\textwidth,page=4]{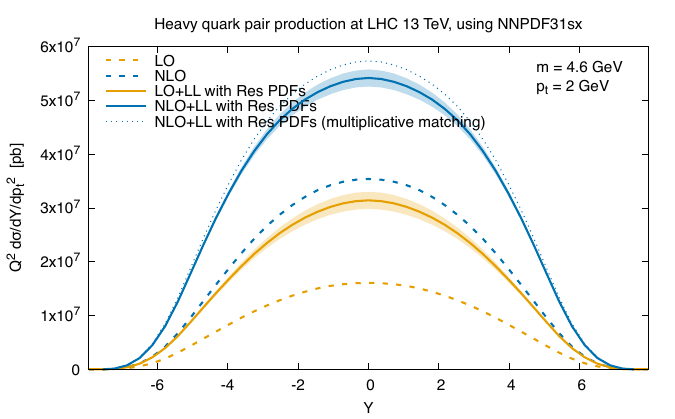}\\
  \includegraphics[width=0.49\textwidth,page=7]{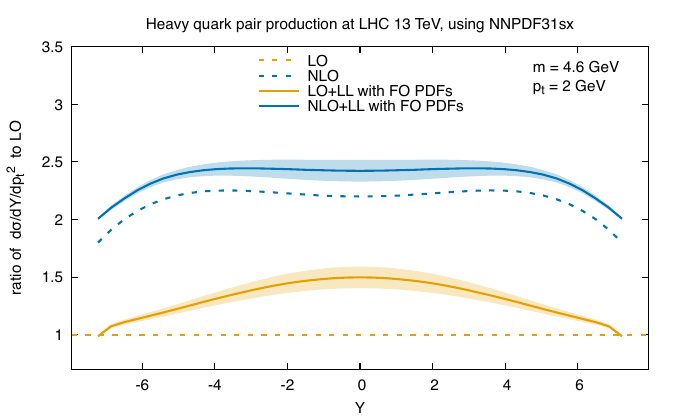}
  \includegraphics[width=0.49\textwidth,page=8]{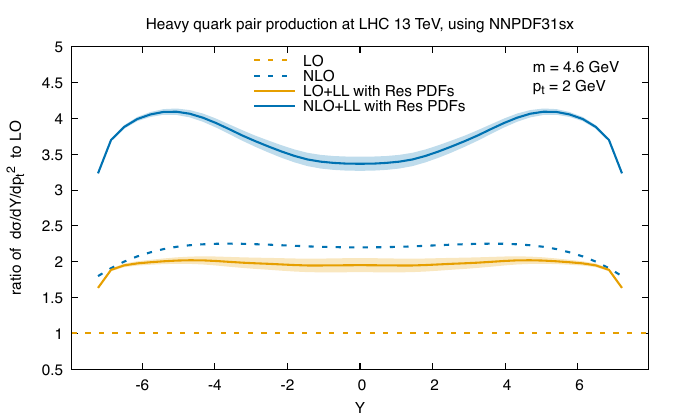}
  \caption{The double differential distribution in rapidity and transverse momentum of the bottom quark,
    plotted as a function of the rapidity for $p_t=2$~GeV, for bottom pair production at LHC $13$~TeV.
    The left plots are obtained using NNPDF31sx at fixed order, while in the right plot the resummed result
    is computed with the resummed PDFs from the same family.
    The uncertainty band represents an estimate of NLL corrections.}
  \label{fig:SQDoubleDiff}
\end{figure}

To appreciate the effect of the resummed contributions on physical cross sections,
we present the differential distributions after convolution with the PDFs in Fig.~\ref{fig:SQDoubleDiff},
considering for definiteness bottom pair production at LHC $13$~TeV.
We use the NNPDF31sx~\cite{Ball:2017otu} PDF set that has been obtained in the context of a study
on the inclusion of small-$x$ resummation in PDF fits.
The advantage of this set is that it provides PDFs consistently obtained with and without the inclusion of
small-$x$ resummation.
In the following, we will use the same fixed-order PDFs to compute both the fixed-order and the resummed result,
in order to emphasise the effect of resummation in the perturbative coefficient.
Also, we provide resummed results obtained with the resummed PDFs, to see how much the resummation in PDFs impacts the cross section.
However, performing a thorough phenomenological study is beyond the scope of this paper and is left to future work.

The plots of Fig.~\ref{fig:SQDoubleDiff} show the double differential distribution in rapidity $Y$ and transverse momentum $p_t$
at various orders (upper plots) and their ratio to the LO (lower plots),
as a function of $Y$ and for fixed $p_t=2$~GeV.
In the left plots, we use the same (fixed-order) PDFs for both fixed-order results and resummed results.
We show the LO cross section in dashed orange and the NLO one in dashed blue.
The latter, obtained from POWHEG-box~\cite{Nason:2004rx,Frixione:2007vw,Alioli:2010xd},
is about twice as large as the LO result, which is partly due to the large value of $\as$ at this low scale.\footnote
{We use $\mur=\muf=m_b$, which is not the standard choice in POWHEG and probably not the best choice
  in terms of stability of the perturbative expansion, but it allows a simple matching with the resummed contribution.}
In solid we plot the LO+LL (orange) and NLO+LL (blue) results.
We observe that resummation is a positive correction at LO, of about 50\% at central rapidity
and decreasing towards the rapidity endpoints.
At NLO, the correction of resummation is still positive, but smaller in size, showing that the
perturbative expansion converges better when resummation is included.
Overall, the NLO+LL result is approximately a 140\% correction over the LO across the whole rapidity range
except towards the endpoints, where it goes down a bit following the analogous behaviour of the NLO.
When the resummed LO+LL and NLO+LL results are computed with resummed PDFs (right plots),
the impact of resummation becomes much larger, as a consequence of the fact that the resummed gluon
is larger than the fixed-order one at small $x$~\cite{Ball:2017otu,Abdolmaleki:2018jln,Bonvini:2019wxf}.
In particular, the NLO+LL curve has a large K-factor at large rapidities,
where the contribution from the gluon at small $x$ is dominant.
This shows that this observable is very sensitive to the PDFs at small $x$,
and it thus represents an important process to give additional constraints to PDF fits,
in agreement with the findings of Refs.~\cite{PROSA:2015yid,Gauld:2015yia,Gauld:2016kpd}.

\begin{figure}[t]
  \centering
  \includegraphics[width=0.49\textwidth,page=1]{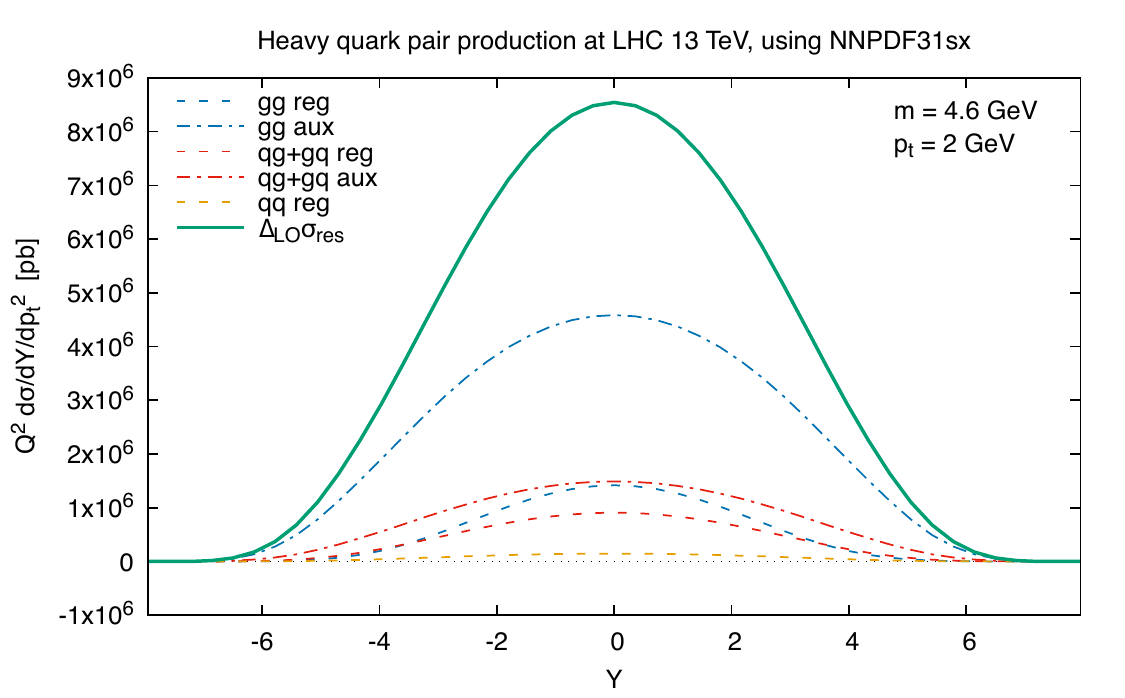}
  \includegraphics[width=0.49\textwidth,page=2]{images/plot_QQbarSQ}
  \caption{Breakdown of the individual contributions to the resummed result from the $gg$, $gq+qg$ and $qq$ channels
    separating the regular and auxiliary parts. The left plot focuses on the resummed contribution to be matched to the LO,
    while the right plot focuses on the resummed contribution to be matched to NLO.
    The results in these plots are obtained using NNPDF31sx with resummation.}
  \label{fig:SQDoubleDiff2}
\end{figure}

It is interesting to understand how the various contributions add up to form the resummed result.
First of all, we stress that the LO cross section is made of two contributions, one in the $gg$ channel
and one in the $q\bar q$ channel. The second one, however, is very small, so the LO curve
is almost entirely given by its $gg$ contribution.
As far as the resummed result is concerned, we not only distinguish between channels but also
between the regular and auxiliary contributions, as given in Eqs.~\eqref{eq:resC_reg_aux}.
The breakdown of the individual resummed contributions to be added to the LO is shown in Fig.~\ref{fig:SQDoubleDiff2} (left).
We observe that the dominant contributions are those coming from the auxiliary part,
both in the $gg$ channel and in the $qg+gq$ channel. The regular contributions are smaller
and localised in a region of central rapidity.
Also, we note a clear hierarchy in the contributions by the various channels,
with the $gg$ dominating over the $qg+gq$, and the $qq$ being very small.
We also stress that the $qg+gq$ channel is symmetric because we plot them together, but each individual
contribution, $qg$ and $gq$, is obviously asymmetric (see Fig.~\ref{fig:SQDoubleDiffPartonic}).
The right plot of Fig.~\ref{fig:SQDoubleDiff2} shows the analogous breakdown
for the resummed contribution to be added to the NLO to construct the NLO+LL result.
The difference here is only in the auxiliary contributions,
as the regular contribution starts at relative $\Ord(\as^2)$ and is thus unaffected when subtracting the expansion at $\Ord(\as)$.
Because of this subtraction, the auxiliary contributions become comparable with the regular ones at mid rapidities,
but they still dominate in the forward region, as expected.

In order to understand the stability of the resummed result, we now discuss its uncertainties.
Because our resummed results are accurate at LL only,
the first uncertainty we consider is the one coming from the unknown subleading logarithmic contributions.
In previous \texttt{HELL} works~\cite{Bonvini:2016wki,Bonvini:2017ogt,Bonvini:2018iwt,Bonvini:2018xvt}
such uncertainty is studied by varying subleading ingredients in the construction
of the resummed anomalous dimension entering the evolution function Eq.~\eqref{eq:UABFht}
in two different ways,\footnote
{One variation is given by a modified way of implementing the resummation of subleading running coupling contributions in the
  anomalous dimension~\cite{Bonvini:2017ogt}. The other variation makes use of what we called LL$^\prime$ anomalous dimension
  introduced in Ref.~\cite{Bonvini:2016wki} in place of the full NLL one, which gives by far the largest contribution to the uncertainty
  (see also Ref.~\cite{Bonvini:2018iwt}).}
and by varying the form of the evolution function itself
by replacing $r(N,\as)$, Eq.~\eqref{eq:UABF} with $\as\beta_0$.
The effect of these three independent ways of varying subleading logarithms in the resummed
result is then added in quadrature to form a representative uncertainty for the final result.
We adopt this procedure here, and we show the resulting uncertainty as a band in Fig.~\ref{fig:SQDoubleDiff}.
While this way of computing the uncertainties may possibly underestimate the actual size of NLL contributions,
it is clear from the plot that the difference between LO+LL and NLO+LL cannot be due
to subleading logarithms only, as it is much larger than their uncertainty.
Therefore, at this scale and value of $\pt$,
contributions that are subleading power at small $x$ are important.
This can be seen also by looking at the difference between
additive matching (our default) and multiplicative matching,\footnote
{We recall that additive matching means that the resummed contribution is added to the fixed order subtracting the doubly counted contributions
  (corresponding to the expansion of the resummed result up to the order at which the fixed order is computed),
  while in the multiplicative matching the fixed order is multiplied by the resummed result divided by its expansion.}
shown as a dotted line in the plot.
The difference between these two curves, being related to the ratio between the
exact NLO and its small-$x$ approximation, also includes the effect of subleading power contributions,
and it is indeed outside the uncertainty band from subleading logarithms.

\begin{figure}[t]
  \centering
  \includegraphics[width=0.49\textwidth,page=6]{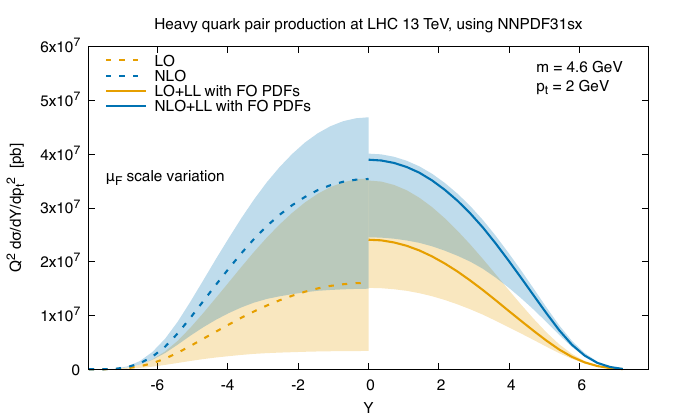}
  \includegraphics[width=0.49\textwidth,page=5]{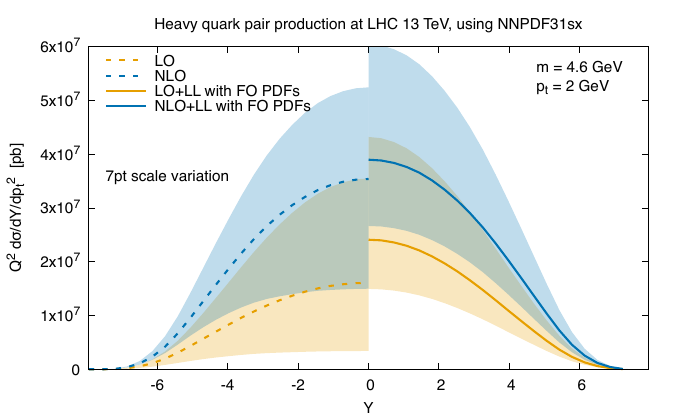}
  \caption{Scale uncertainty for the double differential distribution in rapidity and transverse momentum of the bottom quark,
    plotted as a function of the rapidity for $\pt=2$~GeV, for bottom pair production at LHC $13$~TeV.
    The left plot shows factorization scale uncertainty only, while the right plot shows the standard 7-point uncertainty envelope.}
  \label{fig:SQDoubleDiffScaleUnc}
\end{figure}

In Fig.~\ref{fig:SQDoubleDiffScaleUnc} we also show the scale uncertainty band of our results.
In the left plot, we consider only the factorization scale variation by a canonical factor of 2 up and down,
while in the right plot we construct the envelope of the customary 7-point variation of $\muf$ and $\mur$.
Because the rapidity distribution is symmetric, in each plot we show the fixed-order result for negative rapidity
and the resummed result for positive rapidity, for a better visualisation of the bands.
As far as $\muf$ variation is concerned, we note a clear reduction of the uncertainty
after the inclusion of the resummation,
demonstrating the perturbative stabilisation that small-$x$ resummation allows to achieve.
However, the uncertainty of the resummed result becomes comparable to the one of the fixed order once $\mur$ variations are also taken into account.
This is not surprising, for two reasons.
The first one is that the value of $\as$ varies significantly as $\mur$ changes because the scale of the process is low
(for the same reason, the NLO uncertainty is larger than the LO one).
The other reason is that at LL there are no logarithms of $\mur$ in the resummed result
to compensate for the change in $\as$, as $\mur$ dependence in the resummation starts at NLL.
To see a reduction of the 7-point uncertainty band the resummation should be performed at the currently unknown NLL order.

\begin{figure}[t]
  \centering
  \includegraphics[width=0.69\textwidth,page=3]{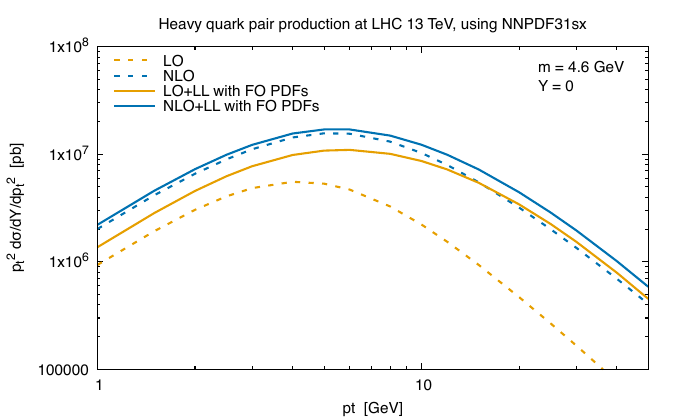}
  \caption{The double differential distribution in rapidity and transverse momentum of the bottom quark,
    plotted as a function of $p_t$ for central rapidity $Y=0$, for bottom pair production at LHC $13$~TeV.}
  \label{fig:SQDoubleDiffPt}
\end{figure}

To conclude the section, we now consider the same double differential rapidity distribution
but as a function of $\pt$ at fixed central rapidity $Y=0$.
This is shown with fixed-order PDFs in Fig.~\ref{fig:SQDoubleDiffPt}.
We observe that going towards large transverse momentum two effects are manifest:
the NLO correction grows, and the impact of resummation on the LO gets larger while
matching resummation to NLO gives a smaller correction.
This suggests that the large NLO contribution at large $\pt$ is dominated by small-$x$ logarithms,
and once these are resummed the perturbative convergence improves significantly.
As the resummation has no direct dependence on the transverse momentum other than in kinematic constraints,
this is just a consequence of the kinematics.
In particular, we suspect that the smaller available phase space at large $\pt$
makes contributions from the low-$x$ region dominant also at central rapidity
(at large rapidity this is expected at any $p_t$).
We plan to investigate this effect further in future phenomenological studies.

\subsection{Results differential in the heavy-quark pair}
\label{sec:pairkin}

In this section we consider the final state to be the heavy-quark pair,
and so focus on the differential distribution in the components of the momentum $q=p+\bar p$
which is the sum of the momenta of the two heavy quarks.
For instance, this choice is appropriate for describing the measurement of a bound state of the heavy quarks,
e.g.\ the $J/\psi$ for $c\bar c$ pairs or the $\Upsilon$ for $b\bar b$ pairs or heavier resonances.
The details of the computation of the partonic off-shell coefficient function are given in App.~\ref{app:pairkin}.
Because at the lowest order the process is effectively a 2 to 1 process,
the differential coefficient function contains delta functions, Eq.~\eqref{eqP:triplediffxs2}.
This implies that the computation of some of the integrals defining the resummed collinear coefficient functions,
as described in section~\ref{sec:channels}, can be carried out analytically,
partly simplifying the numerical implementation.
Explicit expressions are presented in App.~\ref{app:PairCF}.

Note that these simplifications pose some problems in presenting the results.
Indeed, for instance, for the triple differential distribution
the regular coefficient function Eq.~\eqref{eq:resCreg} is an actual function,
while the auxiliary coefficient Eq.~\eqref{eq:resCaux} is a distribution,
making a visual comparison at parton level impossible.
This problem can be overcome by showing the cross section at hadron level only,
after integration with the PDFs.
For definiteness, we consider bottom pair production at LHC $13$~TeV, with bottom mass $m_b=4.6$~GeV,
as done in the previous section.
Similarly, we use the same NNPDF31sx~\cite{Ball:2017otu} PDF set considered before.

\begin{figure}[t]
  \centering
  \includegraphics[width=0.49\textwidth,page=3]{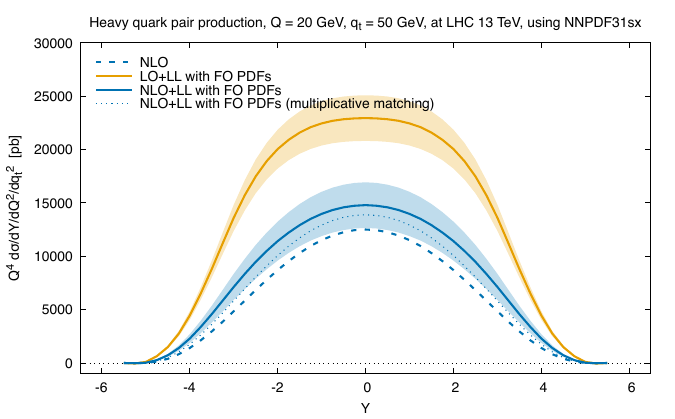}
  \includegraphics[width=0.49\textwidth,page=4]{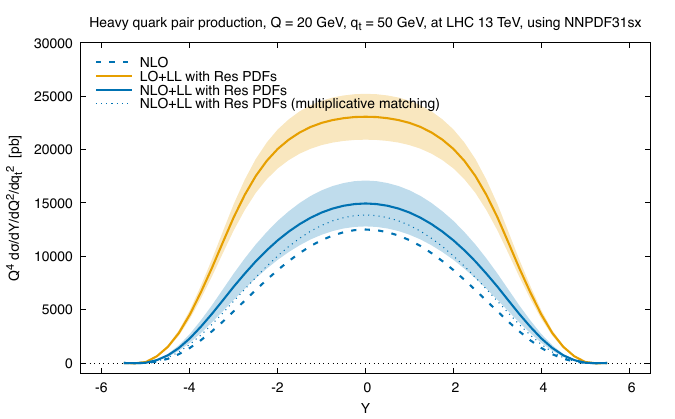}
  \caption{The triple differential distribution in invariant mass, rapidity and transverse momentum of the bottom pair,
    plotted as a function of the rapidity for $Q=20$~GeV and $p_t=50$~GeV, for bottom pair production at LHC $13$~TeV.
    The left plots are obtained using NNPDF31sx at fixed order, while in the right plot the resummed result
    is computed with the resummed PDFs from the same family.
  The uncertainty band represents an estimate for the NLL corrections.}
  \label{fig:TripleDiff}
\end{figure}

In Fig.~\ref{fig:TripleDiff} we show the triple differential distribution,
plotted as a function of the rapidity $Y$ of the pair
and at fixed invariant mass $Q=20$~GeV and fixed transverse momentum $q_t=50$~GeV.
In this case, the LO curve is not present, as it is proportional to $\delta(\qt^2)$,
and so it is zero for any non-zero value of the transverse momentum.
Consequently, we cannot show a ratio plot.
We observe that the NLO (blue dashed curve) is smaller than the LL curve (solid orange),
which is effectively a LO+LL result.
After matching with the NLO, the resummed NLO+LL curve (solid blue) represents
a small positive correction to the NLO result, pointing toward the still larger LL prediction.
This suggests that the inclusion of resummation tends to predict a higher cross section than at NLO,
and possibly leads to a better convergence of the perturbative expansion.
As we did for Fig.~\ref{fig:SQDoubleDiff}, we show on the left the resummed result
computed with the same fixed-order PDFs used for the NLO, while we show on the right panel
the resummed contribution computed using the resummed PDFs.
In this case, the difference between the two options is very mild,
probably due to the larger value of $\tau$ and the larger invariant mass,
showing that this observable is not particularly powerful in constraining the PDFs at small $x$.

\begin{figure}[t]
  \centering
  \includegraphics[width=0.49\textwidth,page=1]{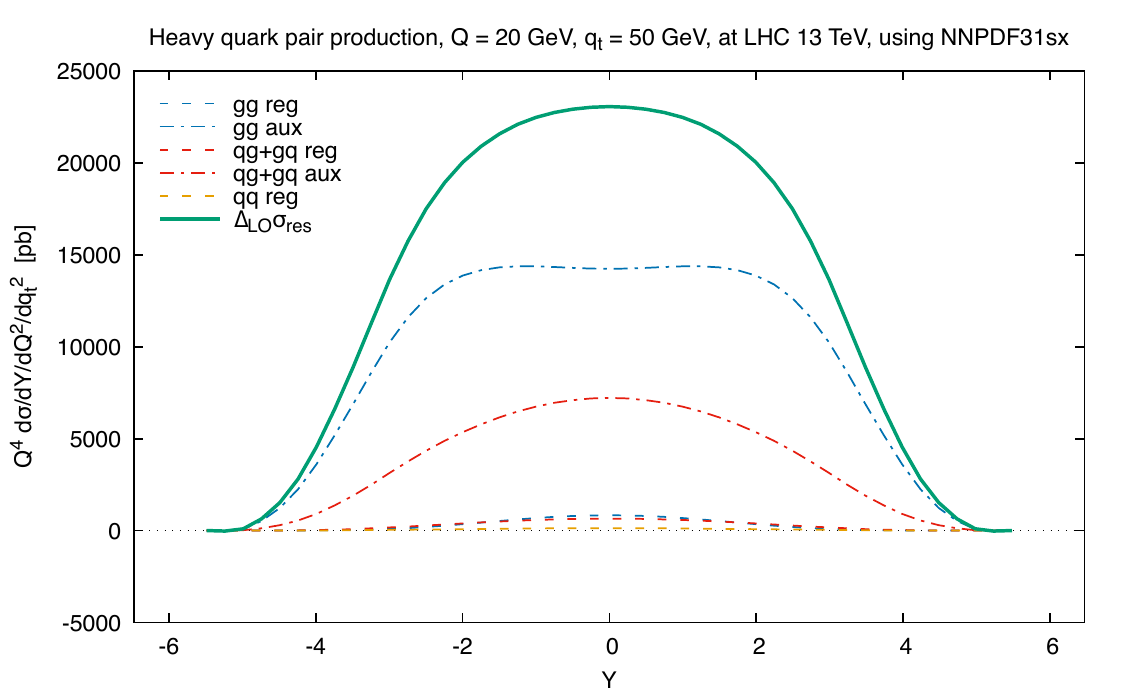}
  \includegraphics[width=0.49\textwidth,page=2]{images/plot_QQbarTripleDiff_qt50}
  \caption{Breakdown of the individual contributions to the resummed triple differential distribution
    in invariant mass, rapidity and transverse momentum of the bottom pair
    from the $gg$, $gq+qg$ and $qq$ channels separating the regular and auxiliary parts.
    The left plot focuses on the resummed contribution to be matched to the LO,
    while the right plot focuses on the resummed contribution to be matched to NLO.
    The results in these plots are obtained using NNPDF31sx with resummation at LHC 13 TeV,
    as a function of the rapidity, for invariant mass $Q=20$~GeV and for transverse momentum
    $q_t=50$~GeV.}
  \label{fig:TripleDiff2}
\end{figure}

Similarly to Fig.~\ref{fig:SQDoubleDiff2}, we also show the breakdown of the individual contributions to the cross section
in Fig.~\ref{fig:TripleDiff2}.
When matching to LO (left plots) we note a pattern similar to what was observed for the single-quark distributions
in the previous section. Namely, the auxiliary term dominates over the regular contribution,
and the $gg$ channel is larger than the $qg+gq$, in turn larger than the $qq$ channel.
The resummed contribution is positive, consistent with the fact that this pure LL distribution is effectively a LO+LL result.
When subtracting the $\Ord(\as)$ expansion to match the resummation to NLO (right plots)
we find a smaller contribution from resummation.
Again, we note that the auxiliary contributions are now comparable in size with the regular ones at mid rapidities,
but they keep giving a much larger contribution at large rapidities.

\begin{figure}[t]
  \centering
  \includegraphics[width=0.49\textwidth,page=6]{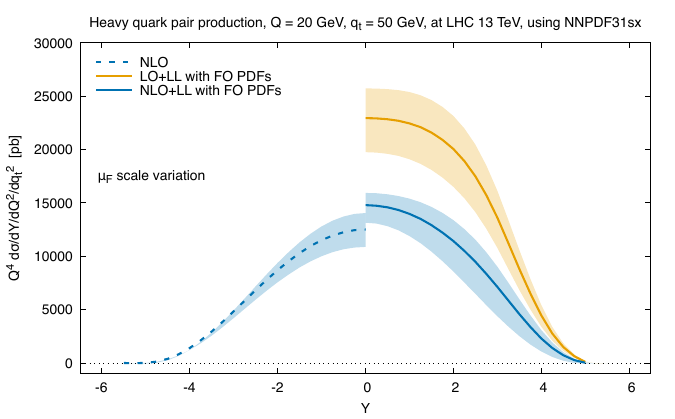}
  \includegraphics[width=0.49\textwidth,page=5]{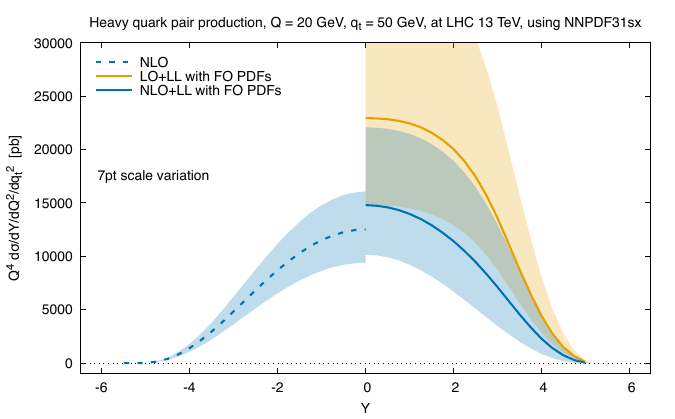}
  \caption{Scale uncertainty for the triple differential distribution in rapidity and transverse momentum of the bottom quark pair,
    plotted as a function of the rapidity for $\pt=50$~GeV, for bottom pair production at LHC $13$~TeV.
    The left plot shows factorization scale uncertainty only, while the right plot shows the standard 7-point uncertainty envelope.}
  \label{fig:PairTripleDiffScaleUnc}
\end{figure}

We conclude the section by briefly commenting on the uncertainties.
In Fig.~\ref{fig:TripleDiff} the resummed curves are supplemented with an uncertainty band computed as discussed in the previous section
to estimate the impact of subleading logarithmic contributions.
This uncertainty is relatively larger than in the case of single-quark kinematics,
but still it cannot account for the full difference between LO+LL and NLO+LL,
which thus gets significant contributions from non-small-$x$ effects.
The use of multiplicative matching at NLO+LL, probing some subleading power contributions,
differs from the additive matching by an amount that is comparable with the uncertainty band from subleading logarithms.
Moving to scale variations, we show in Fig.~\ref{fig:PairTripleDiffScaleUnc}
$\mu_F$ variations on the left plot and a full 7-point variation on the right plot.
Considerations similar to what we have done for the single-quark kinematics apply.
We limit ourselves to observe that in both plots there is a visible reduction moving from LO+LL to NLO+LL,
again hinting at a stabilisation of the perturbative expansion once resummation is included.

Further studies of different distributions and different kinematic configurations,
relevant for phenomenological applications, are beyond the scope of this more theoretical paper
and are left to future work.

\section{Conclusions}
\label{sec:conclusions}

In this paper we have extended the \texttt{HELL} formalism for the small-$x$ resummation of physical observables
to differential distributions at LL.
We have obtained resummed formulae for differential partonic coefficient functions
which are valid for any process that is gluon-gluon initiated at LO.
The application of the formalism to other kinds of processes requires the
treatment of collinear subtractions to all orders at small $x$,
whose extension at differential level is left to future work~\cite{DYsx}.

With respect to previous implementations of small-$x$ resummation,
we no longer perform an approximation, valid at LL, where the off-shell coefficient function
was computed at $N=0$ in Mellin space.
This approximation simplifies the resummation of inclusive cross sections where such Mellin
transform could be computed analytically.
Here, because in general we are not able to compute this Mellin transform analytically at differential level,
adopting such approximation would not lead to any simplification.
Rather, it would break the kinematic limits of the observables,
which is clearly undesirable.

We have considered heavy-quark pair production at proton-proton colliders as a representative application of our results.
We have resummed distributions differential both in the momentum of a single heavy quark and
in the sum of the momenta of both heavy quarks (momentum of the pair).
The selection of numerical results presented serves as a demonstration that the methodology works
and that it can be used for phenomenology.
They also show that the impact of small-$x$ resummation for these observables is significant,
as we expect from the low-$x$ values that heavy quark pair production can reach at LHC.
However, the results presented in this work do not represent a full phenomenological study,
that would also require the description of the hadronisation of the heavy quarks in order
to compare with the data. Such a phenomenological study, with the goal to include the process
in a PDF fit to improve the PDF quality at low $x$, is left to future work.

The new version of the \texttt{HELL} code, that implements the resummation of
heavy quark pair production at differential level, is available at the url
\begin{center}
  \href{http://www.roma1.infn.it/~bonvini/hell}{\tt www.roma1.infn.it/$\sim$bonvini/hell}
\end{center}
The preparation of tables for quick interpolation, needed for phenomenological applications,
requires some time and leads to a large amount of data, because of the dependence on many kinematic variables.
Therefore, rather than providing general tables within the code (as previously done for DIS and Higgs),
we only provide scripts for the generation of such tables, which can then be produced and used
directly by the user focussing only on the kinematics of interest.

\acknowledgments
{
  We thank Simone Marzani, Luca Rottoli and Francesco Giovanni Celiberto for several useful discussions
  and for comments on the manuscript.
  The work of MB was supported by the Marie Sk\l{}odowska Curie grant HiPPiE@LHC under the agreement n.~746159.
}

\appendix
\section{The off-shell coefficient function}
\label{app:XS}

In this Appendix we give all the details for the computation of the off-shell coefficient function
for heavy quark pair production at proton-proton colliders.
The partonic process at the lowest order, relevant for LL resummation, is
\begin{align}
g^*\(k_1\)+g^*\(k_2\) \rightarrow Q\(p\)+\bar Q\(\bar p\).
\end{align}
where $Q$ and $\bar Q$ are the two heavy quarks of mass $m$.
We parametrize the momenta as
\begin{subequations}
\begin{align}
k_1 &= x_1 P_1 + \kvec_1 \\
k_2 &= x_2 P_2 + \kvec_2 \\
p &= z_1 x_1 P_1 + z_2x_2 P_2 + \pvec \\
\bar p&= (1-z_1)x_1 P_1 + (1-z_2) x_2 P_2 + \kvec_1 + \kvec_2 -\pvec,
\end{align}
\end{subequations}
where, in the collider center-of-mass frame, the protons momenta are
\beq
P_1 = \frac{\sqrt{s}}2(1,0,0,1),\qquad
P_2 = \frac{\sqrt{s}}2(1,0,0,-1).
\eeq
In these definitions we have already used momentum conservation, and we have made a choice of reference frame.
There are 7 initial-state parameters ($s,x_1,x_2,\kvec_1,\kvec_2$) and 4 final-state parameters ($z_1,z_2,\pvec$).
Note however that using the on-shell condition for the final-state quarks we can constrain one of the final-state parameters.
Indeed, there are two on-shell conditions,
\begin{subequations}\label{eq:onshellppbar1}
\begin{align}
m^2 = p^2 &= z_1z_2x_1x_2s-|\pvec|^2 \\
m^2 =\bar p^2 &= (1-z_1)(1-z_2)x_1x_2s-|\kvec_1+\kvec_2-\pvec|^2,
\end{align}
\end{subequations}
setting the squared momenta $p^2$ and $\bar p^2$ to the \emph{same} mass $m^2$.
Therefore, only three of the four final state parameters are independent.

The partonic off-shell coefficient function is computed in the ``partonic'' reference frame,
that corresponds to the partonic center-of-mass frame if the two gluons were on shell, namely if $\kvec_1=\kvec_2=0$.
In other words, the partonic frame is related to the collider frame by a longitudinal boost of rapidity
\beq
\bar\eta= \frac12\log\frac{x_1}{x_2}.
\eeq
In this frame, the partonic coefficient can only depend on $x_1$, $x_2$ and $s$ through the product $x_1x_2s$.
Moreover, because we assume unpolarized protons, an overall azimutal angle is irrelevant.
Thus the coefficient can only depend on 4 out of the 7 initial-state parameters.
We choose them to be
\begin{subequations}\label{eq:initialstatevar}
\begin{align}
z &\equiv \frac{Q^2}{x_1x_2s}\\
\xi_1 &\equiv \frac{|\kvec_1|^2}{Q^2} \\
\xi_2 &\equiv \frac{|\kvec_2|^2}{Q^2} \\
\varphi &\equiv \text{angle between $\kvec_1$ and $\kvec_2$}.
\end{align}
\end{subequations}
Here, $Q^2$ is ``the hard scale'', whose value depends on the final state we want to look at.
We set $Q^2=q^2$, where $q$ is the final state momentum with respect to which we want to be differential.
In particular, if we want to study the kinematics of the heavy-quark pair, then $q=p+\bar p$ and $Q^2$ is the squared invariant mass of the pair,
while for the single heavy quark then $q=p$ and $Q^2=m^2$ is the mass squared of the quark itself.

Before discussing each of these cases in turn,
we note that $x_1x_2s$ is \emph{not} $\hat s=(k_1+k_2)^2=x_1x_2s-|\kvec_1+\kvec_2|^2$, because of the transverse component of the gluons;
we may call it the ``longitudinal part'' of $\hat s$ (meaning the contribution to $\hat s$ due to the longitudinal part of the gluon momenta).
The full partonic center-of-mass energy $\hat s$ can be written as
\beq\label{eq:shatnewvar}
\hat s = Q^2\[\frac1z-\xi_1 -\xi_2-2\sqrt{\xi_1\xi_2}\cos\varphi\]
\eeq
in terms of the new variables, which reduces to the usual expression $\hat s = Q^2/z$ when the gluons are on shell.

\subsection{Kinematics for the single quark}
\label{app:singlekin}

Here we consider the differential distribution in the kinematics of one of the final-state heavy quarks.
For definiteness, we consider the heavy quark of momentum $p$, but since the process is symmetric
the results will equally apply also to the antiquark with momentum $\bar p$.
We introduce the variables
\begin{subequations}\label{eq:finalstatevars}
\begin{align}
Q^2 &\equiv p^2 = z_1z_2x_1x_2s-|\pvec|^2 = m^2 \\
\eta &\equiv \frac12\log\frac{p^0+p^3}{p^0-p^3}-\bar\eta= \frac12\log\frac{z_1}{z_2} \\
\pthat^2 &\equiv \frac{\pvec^2}{Q^2} = \frac{\pvec^2}{m^2} \\
\vartheta &= \text{angle between $\pvec$ and $\kvec_1+\kvec_2$}.
\end{align}
\end{subequations}
Because $Q^2=m^2$ is fixed, the most differential distribution we are interested in is ($\pt^2=\pthat^2Q^2$)
\beq\label{eq:triplediffxs}
\frac{\dd{\cal C}}{\dd \eta\, \dd\pt^2}(z,\xi_1,\xi_2,m^2, \eta,\pthat^2),
\eeq
which is integrated over $\vartheta$ and averaged over $\varphi$.
Note that from now on we are omitting the argument $\as$ from the off-shell distribution
as we are interested in the lowest order result only.

Let us consider the phase space. The two-body phase space is given by
\begin{align}
  \dd\phi_2(k_1+k_2;p,\bar p)
  &= \theta(\hat s-4m^2)\frac{\dd^4p}{\left(2\pi\right)^3}\frac{\dd^4\bar p}{\left(2\pi\right)^3}
    \delta\left(p^2-m^2\right)\delta\left(\bar p^2-m^2\right)\left(2\pi\right)^4\delta^{(4)}\left(k_1+k_2-p-\bar p\right)
    \theta(p^0) \theta(\bar p^0) \nonumber \\
  &=\theta(\hat s-4m^2)\frac{\dd^4p}{4\pi^2}\delta\(p^2-m^2\)\delta\left((k_1+k_2-p)^2-m^2\right)
    \theta(p^0) \theta(k_1^0+k_2^0-p^0)
\end{align}
with $\hat s = (k_1+k_2)^2$.
We need to express this phase space in terms of the new variables.
The variable $\hat s$ is given in Eq.~\eqref{eq:shatnewvar},
the integration element can be written as
\beq
\dd^4p = \frac{Q^2}4 \,\dd Q^2\,\dd \eta\,\dd\pthat^2\,\dd\vartheta,
\eeq
and the antiquark momentum squared is
\begin{align}\label{eq:pp2}
\bar p^2 &= (k_1+k_2-p)^2 \nonumber\\
&= (1-z_1)(1-z_2)x_1x_2s-|\kvec_1+\kvec_2-\pvec|^2 \nonumber\\
&= Q^2\bigg[1+\frac1z-\sqrt{\frac{1+\pthat^2}{z}}(e^{\eta}+e^{-\eta})-\xi_1-\xi_2-2\sqrt{\xi_1\xi_2}\cos\varphi
\nonumber\\ &\qquad\qquad
+2\sqrt{\(\xi_1+\xi_2+2\sqrt{\xi_1\xi_2}\cos\varphi\)\pthat^2}\cos\vartheta\bigg],
\end{align}
where we have used the inverse relations
\beq\label{eq:z12pty}
z_1 = \sqrt{z(1+\pthat^2)}e^{\eta},\qquad
z_2 = \sqrt{z(1+\pthat^2)}e^{-\eta}.
\eeq
The conditions imposed by the two theta functions in the energies translate easily into conditions on $z_1$ and $z_2$
that depend on $x_1$ and $x_2$, namely $z_1x_1+z_2x_2\geq0$ and $(1-z_1)x_1+(1-z_2)x_2\geq0$.
From the on-shell conditions Eq.~\eqref{eq:onshellppbar1} we also know that $z_1z_2x_1x_2\geq0$ and $(1-z_1)(1-z_2)x_1x_2\geq0$.
Because $x_1$ and $x_2$ are positive, it follows that $z_1$ and $z_2$ satisfy the conditions $0\leq z_{1,2}\leq1$,
that translate into
\beq\label{eq:z12condition}
z(1+\pthat^2)\leq e^{-2\abs{\eta}}.
\eeq
After the trivial integration over $Q^2$, the phase space can thus be recast as
\begin{align}
\dd\phi_2
  &= \theta\(\frac1z-\xi_1-\xi_2-2\sqrt{\xi_1\xi_2}\cos\varphi-4\)\, \theta\(\frac1z-(1+\pthat^2)e^{2\abs{\eta}}\)\,
    \frac1{16\pi^2}\,\dd \eta\,\dd\pthat^2\,\dd\vartheta\\
  &\times\delta\(\frac1z-\sqrt{\frac{1+\pthat^2}{z}}(e^{\eta}+e^{-\eta})-\xi_1-\xi_2-2\sqrt{\xi_1\xi_2}\cos\varphi+2\sqrt{\(\xi_1+\xi_2+2\sqrt{\xi_1\xi_2}\cos\varphi\)\pthat^2}\cos\vartheta\). \nonumber
\end{align}
To simplify the notation, we introduce the function
\beq
\xi(\xi_1,\xi_2,\varphi) = \xi_1+\xi_2+2\sqrt{\xi_1\xi_2}\cos\varphi = |\kvec_1+\kvec_2|^2,
\eeq
and simply write $\xi$ without arguments for short.
Putting everything together we have
\begin{align}
\frac{Q^2\dd{\cal C}}{\dd \eta\, \dd\pt^2}(z,\xi_1,\xi_2,m^2, \eta,\pthat^2)
&= \sigma_0\frac12 \int_0^{2\pi}\frac{\dd\varphi}{2\pi} \int \frac{\dd\phi_2}{\dd \eta\, \dd\pthat^2}
\;|{\cal M}|^2 \\
&=\frac{\sigma_0}{32\pi^2} \int_0^{2\pi}\frac{\dd\varphi}{2\pi}\, \theta\(\frac1{z}-\xi-4\)\, \theta\(\frac1z-(1+\pthat^2)e^{2\abs{\eta}}\) \int_0^{2\pi}\dd\vartheta
\;|{\cal M}|^2 \nonumber\\
&\quad \times\delta\(\frac1z-\sqrt{\frac{1+\pthat^2}{z}}(e^{\eta}+e^{-\eta})-\xi+2\sqrt{\xi\pthat^2}\cos\vartheta\),\nonumber
\end{align}
where in the first line $1/2$ is the flux factor, $\sigma_0=16\pi^2\as^2/Q^2$ and the $1/2\pi$ comes from the average over $\varphi$.
The matrix element squared $|{\cal M}|^2$ is given in Appendix~\ref{app:ME}.

It is most convenient to use the $\delta$ function to integrate over $\vartheta$,
as all other variables appear at least quadratically.
The fact that $|\cos\vartheta|\leq1$ produces the constraint
\begin{align}
\left|\frac1z-\sqrt{\frac{1+\pthat^2}{z}}(e^{\eta}+e^{-\eta})-\xi\right|
&\leq 2\sqrt{\xi\pthat^2}.
\end{align}
We then get
\begin{align}\label{eq:dCoffFinSQ}
\frac{Q^2\dd{\cal C}}{\dd \eta\, \dd\pt^2}(z,\xi_1,\xi_2,m^2, \eta,\pthat^2)
  &=\frac{\sigma_0}{32\pi^2} \int_0^{2\pi}\frac{\dd\varphi}{2\pi}\,
    \theta\(\frac1{z}-\xi-4\)
  \, \theta\(\frac1z-(1+\pthat^2)e^{2\abs{\eta}}\)
\nonumber\\
&\qquad \times\theta\(2\sqrt{\xi\pthat^2}-\abs{\frac1{z}-\sqrt{\frac{1+\pthat^2}{z}}(e^{\eta}+e^{-\eta})-\xi}\)
\nonumber\\
&\qquad \times\frac{|{\cal M}|^2_{\vartheta=\bar\vartheta}+|{\cal M}|^2_{\vartheta=2\pi-\bar\vartheta}}{\sqrt{4\xi\pthat^2-\(\frac1{z}-\sqrt{\frac{1+\pthat^2}{z}}(e^{\eta}+e^{-\eta})-\xi\)^2}} 
,\\
\bar\vartheta &= \cos^{-1}\frac{\xi-\frac1z+\sqrt{\frac{1+\pthat^2}{z}}(e^{\eta}+e^{-\eta})}{2\sqrt{\xi\pthat^2}},\qquad 0\leq\vartheta\leq\pi.
\end{align}
The theta functions in Eq.~\eqref{eq:dCoffFinSQ} may prove troublesome from a numerical point of view.
Indeed, if used as ``if'' conditions that set the integrand to zero when the theta functions are zero,
the numerical integration may become inaccurate.
It is much more convenient to translate them into integration limits of some variable.
To do so, we define
\beq
X = \frac1{\sqrt{z}} \geq 1
\eeq
so that the constraint imposed by the three theta functions become
\begin{subequations}
\begin{align}
  &X\geq\sqrt{4+\xi},\\
  &X\geq\sqrt{1+\pthat^2}e^{\abs{\eta}},\label{eq:condition21}\\
  &-2\sqrt{\xi}\pthat\leq -X^2+2BX+\xi\leq 2\sqrt{\xi}\pthat, \qquad B \equiv \sqrt{1+\pthat^2}\cosh \eta \geq 1.
\end{align}
\end{subequations}
Focussing on $\xi$, we may write
\begin{subequations}
\begin{align}
  &\xi\leq X^2-4, \label{eq:condition11}\\
  &\xi+2\pthat\sqrt{\xi} + 2BX-X^2\geq0, \label{eq:condition12}\\
  &\xi-2\pthat\sqrt{\xi} + 2BX-X^2\leq0. \label{eq:condition13}
\end{align}
\end{subequations}
The functions $\xi\pm 2\pthat\sqrt{\xi} + 2BX-X^2$ represent two parabolae in $\sqrt{\xi}$ with centers (minima) in $\sqrt{\xi}=\mp\pthat$,
at which they both equal $-\pthat^2+2BX-X^2$. If this value is positive, there is no solution to the system,
so we have the condition
\beq\label{eq:condition22}
\pthat^2-2BX+X^2 \geq0
\eeq
that represents an equation for the other variables to be taken into account later, together with Eq.~\eqref{eq:condition21}.
Under this condition, the solution of the inequalities Eq.~\eqref{eq:condition12}, \eqref{eq:condition13} is the region
between the two right solution of the second inequality and the largest between the right solution of the first and the left solution of the second,
which are identical but have opposite sign. Thus we get
\beq\label{eq:conditionxifinal}
\boxed{\left|\pthat-\sqrt{\pthat^2-2BX+X^2}\right| \leq \sqrt{\xi} \leq \pthat+\sqrt{\pthat^2-2BX+X^2}.}
\eeq
The other condition Eq.~\eqref{eq:condition11} is always
automatically satisfied. Indeed, we can prove that
\beq
\sqrt{X^2-4} \geq \pthat+\sqrt{\pthat^2-2BX+X^2}
\eeq
for all meaningful values of $X$ (namely values for which the square roots are real).
Indeed this condition can be manipulated to
\beq
(B^2-\pthat^2)X^2-4BX+4(1+\pthat^2)\geq0,
\eeq
which is always satisfied because the minimum of the quadratic function, located at $X=2B/(B^2-\pthat^2)$, is always non-negative.
Indeed the minimum is proportional to $B^2-1-\pthat^2$ which is non-negative because $B^2\geq 1+\pthat^2$.
Therefore, Eq.~\eqref{eq:conditionxifinal} is the complete condition on $\xi$.

We now focus on the other variables, that must satisfy the inequalities
Eq.~\eqref{eq:condition21} and \eqref{eq:condition22}.
Let us focus on Eq.~\eqref{eq:condition22}, solving it for $X$. The parabola $X^2-2BX+\pthat^2$ has a minimum in $X=B$
where it equals $\pthat^2-B^2$. This is always negative, as by construction $B^2\geq1+\pthat^2>\pthat^2$. Therefore,
there are two separate solutions, $X\geq B+\sqrt{B^2-\pthat^2}$ and $X\leq B-\sqrt{B^2-\pthat^2}$.
However, since we always have
\beq
\sqrt{1+\pthat^2}e^{\abs{\eta}} \geq B,
\eeq
the second solution is not compatible with Eq.~\eqref{eq:condition21}, and it is therefore forbidden.
We are thus left with the condition
\begin{align}\label{eq:condition22new}
  \boxed{X\geq B+\sqrt{B^2-\pthat^2}} \color{gray}\geq B+1\geq2\color{black},
\end{align}
together with Eq.~\eqref{eq:condition21}.
We can show that Eq.~\eqref{eq:condition21} is always compatible with Eq.~\eqref{eq:condition22new}.
Indeed the inequality
\beq
B+\sqrt{B^2-\pthat^2} \geq \sqrt{1+\pthat^2}e^{\abs{\eta}}
\eeq
holds because we can manipulate it into
\beq
\sqrt{B^2-\pthat^2} \geq \sqrt{1+\pthat^2}\(e^{\abs{\eta}}-\cosh\eta\) = \sqrt{1+\pthat^2}\sinh\abs{\eta}
\eeq
and then, squaring both sides (which are both positive) and rearranging,
\beq
(1+\pthat^2)\(\cosh^2\abs{\eta}-\sinh^2\abs{\eta}\)-\pthat^2\geq0 \qquad\Rightarrow\qquad 1+\pthat^2-\pthat^2\geq0
\eeq
which is clearly true. In conclusion, $X$ satisfies only the inequality Eq.~\eqref{eq:condition22new}
which automatically encodes all the others.

It is useful to mention also the conditions on the kinematic limits of the on-shell resummed coefficient,
as well as on the integration variables defining the resummed result.
From Eq.~\eqref{eq:resCreg}, recalling that the first argument of the evolution function
is a momentum fraction and is thus smaller than 1, we obtain the condition $x/z\leq e^{-2|\bar\eta|}$.
Similarly, from Eq.~\eqref{eq:z12condition} we also have $z(1+\pthat^2)\leq e^{-2|\eta|}$.
From the product of the two inequalities, we obtain the condition
\beq
A^2\equiv x(1+\pthat^2) \leq e^{-2|\eta|-2|\bar\eta|} \leq e^{-2|\eta+\bar\eta|} = e^{-2|y|}.
\eeq
The condition $A\leq e^{-|y|}$, with $A\equiv\sqrt{x(1+\pthat^2)}$, represents a constraint on the arguments
of the on-shell coefficient function. However, this is not the most stringent one.
Indeed, looking at the first inequality, we can derive the integration range of $\bar\eta$, which is given by
\beq
\frac{Ae^y-x\pthat^2}{1-Ae^{-y}} \leq e^{2\bar\eta} \leq \frac {1-Ae^y}{Ae^{-y}-x\pthat^2}.
\eeq
For this range to be non-trivial, the upper limit must be larger than the lower limit, leading to
the condition
\beq
e^{|y|}\leq\frac{1+x\pthat^2}{2A}+\sqrt{\frac{(1+x\pthat^2)^2}{4A^2}-1}
\eeq
which is smaller than 1/A in the region where the square root exists,
given by the condition $\pthat^2\leq\frac{1-2\sqrt x}{x}$ or, equivalently,
\beq
x\leq\(\frac{\sqrt{1+\pthat^2}-1}{\pthat^2}\)^2\leq\frac14.
\eeq

To conclude, we recall that the matrix element squared that we will present in appendix~\ref{app:ME}
must be expressed in terms of the variables defined here.
To achieve this, we need to express $z_1,z_2$ in terms of $\pthat, \eta$ through Eq.~\eqref{eq:z12pty},
and to write the product $\kvec_2\cdot\pvec$ appearing in Eqs.~\eqref{eq:T} and \eqref{eq:U} as
\beq
\frac{\kvec_2\cdot\pvec}{Q^2} = \sqrt{\xi_2\pthat^2}\cos(\vartheta+\varphi'),
\eeq
where $\varphi'$ is the angle of $\kvec_1+\kvec_2$ with respect to $\kvec_2$,
which can be computed from the cartesian representation (aligning the $x$ axis along $\kvec_2$)
\begin{align}
\qvec = \dvec{|\qvec|\cos\varphi'}{|\qvec|\sin\varphi'}
= \dvec{|\kvec_2|+|\kvec_1|\cos\varphi}{|\kvec_1|\sin\varphi}
\end{align}
leading to
\begin{align}
\sin\varphi' &= \frac{\sqrt{\xi_1}\sin\varphi}{\sqrt{\qthat^2}} &
\cos\varphi' &= \frac{\sqrt{\xi_2}+\sqrt{\xi_1}\cos\varphi}{\sqrt{\qthat^2}},
\end{align}
which gives the result
\beq\label{eq:varphi'}
\varphi' =
\begin{cases}
  \cos^{-1}\(\frac{\sqrt{\xi_2}+\sqrt{\xi_1}\cos\varphi}{\sqrt{\qthat^2}}\) & \text{if }\sin\varphi\geq0 \\
  2\pi-\cos^{-1}\(\frac{\sqrt{\xi_2}+\sqrt{\xi_1}\cos\varphi}{\sqrt{\qthat^2}}\) & \text{if }\sin\varphi<0.
\end{cases}
\eeq

\subsection{Kinematics for the pair}
\label{app:pairkin}

We now consider the heavy-quark pair as a fictitious intermediate state, with momentum
\begin{align}
q &\equiv p+\bar p \nonumber\\
  &\equiv \alpha_1x_1 P_1 + \alpha_2 x_2 P_2 + \qvec &&\text{\color{gray}(generic parametrization)}\nonumber\\
  &= k_1+k_2 &&\text{\color{gray}(momentum conservation)}\nonumber\\
  &= x_1 P_1 + x_2 P_2 + \kvec_1+ \kvec_2,
\end{align}
where by momentum conservation $\alpha_1=\alpha_2=1$ and $\qvec = \kvec_1+\kvec_2$.
For this intermediate state, we introduce the variables\footnote
{Note that we are using the same names $Q^2$ and $\eta$ that we used for the single quark kinematics, now referring to another momentum.}
\begin{subequations}\label{eqP:finalstatevars}
  \begin{align}
    Q^2 &\equiv q^2 = \alpha_1\alpha_2x_1x_2s-|\qvec|^2 = x_1x_2s-|\kvec_1+\kvec_2|^2 = (k_1+k_2)^2 \equiv \hat s \\
    \eta &\equiv \frac12\log\frac{q^0+q^3}{q^0-q^3}-\bar\eta =\frac12\log\frac{\alpha_1}{\alpha_2}=0
        \qquad\text{\color{gray}(rapidity of $q$ in the partonic frame)} \\
    \qthat^2 &\equiv \frac{\qvec^2}{Q^2}= \frac{|\kvec_1+\kvec_2|^2}{Q^2} \\
    \thetapair &\equiv \text{angle between $\qvec$ and $\kvec_1+\kvec_2$} = 0.
  \end{align}
\end{subequations}
Our goal is to compute the parton-level off-shell coefficient function ($\qt^2=\qthat^2Q^2$)
\beq\label{eqP:triplediffxs}
\frac{\dd{\cal C}}{\dd Q^2\,\dd \eta\, \dd\qt^2}(z,\xi_1,\xi_2,Q^2,\eta,\qthat^2),
\eeq
which is integrated over $\thetapair$ and averaged over $\varphi$.

Let us consider the phase space.
The two-body phase space of the two final state heavy quarks can be factorized
into the phase space of the pair and its ``decay'' as
\begin{align}\label{eqP:PS}
\dd\phi_2(k_1+k_2;p,\bar p) &= \theta(\hat s-4m^2)\int_{4m^2}^{\hat s}\frac{\dd q^2}{2\pi}\, \dd\phi_1(k_1+k_2;q) \, \dd\phi_2(q;p,\bar p)
\end{align}
where
\begin{align}
\dd\phi_1(k_1+k_2;q)&= \frac{\dd^4q}{\(2\pi\)^3}\delta\(q^2-\hat s\)\(2\pi\)^4\delta^{(4)}\(k_1+k_2-q\) \nonumber\\
&=2\pi\, \dd^4q\, \delta\(q^2-\hat s\) \delta^{(4)}\(k_1+k_2-q\),
\end{align}
with $\hat s=(k_1+k_2)^2$ the invariant mass of the pair, and
\begin{align}\label{eq:twobody}
\dd\phi_2(q;p,\bar p)&= \frac{\dd^4p}{\(2\pi\)^3}\frac{\dd^4\bar p}{\(2\pi\)^3}\delta\(p^2-m^2\)\delta\(\bar p^2-m^2\)\(2\pi\)^4\delta^{(4)}\(q-p-\bar p\) \theta(p^0) \theta(\bar p^0).
\end{align}
The full phase space Eq.~\eqref{eqP:PS} can be simplified using the delta function
of the one-body phase space to perform the $q^2$ integral, giving
\begin{align}
  \dd\phi_2(k_1+k_2;p,\bar p)
  &= \theta(\hat s-4m^2)\, \dd^4q \,\delta^{(4)}\(k_1+k_2-q\)\, \dd\phi_2(q;p,\bar p)\nonumber\\
  &= \theta(\hat s-4m^2)\, \dd Q^2\, \dd \eta\, \dd\qthat^2\, \dd\thetapair\, \dd\phi_2(q;p,\bar p)\nonumber\\
  &\qquad\times\delta\(Q^2-\hat s\)\, \delta(\eta)\, \delta\(\qthat^2-\xi_1-\xi_2-2\sqrt{\xi_1\xi_2}\cos\varphi\)\, \delta(\thetapair) \nonumber\\
  &= \theta(Q^2-4m^2)\, \dd Q^2\, \dd \eta\, \dd\qthat^2\, \dd\thetapair\, \dd\phi_2(q;p,\bar p)\nonumber\\
  &\qquad\times\frac1{Q^2}\delta\(1+\qthat^2-\frac1z\)\, \delta(\eta)\, \delta\(\qthat^2-\xi_1-\xi_2-2\sqrt{\xi_1\xi_2}\cos\varphi\)\, \delta(\thetapair),
\end{align}
where we have rewritten $\dd^4q$, the delta function and $\hat s$ in terms of the new variables.

The two-body phase space can be used to integrate the matrix element and remove the ``internal'' degrees of freedom of the pair,
while the one-body phase space can be used to obtain the desired differential observable.
Thus, we immediately find the relation
\begin{align}\label{eqP:triplediffxs2}
  \frac{Q^4\,\dd{\cal C}}{\dd Q^2\,\dd \eta\, \dd\qt^2\, \dd\varphi}(z,\xi_1,\xi_2,Q^2,\eta,\qthat^2,\varphi)
  &= \frac{\dd{\cal C}}{\dd\varphi}(z,\xi_1,\xi_2,Q^2,\varphi)\nonumber\\
  &\times\delta\(1+\qthat^2-\frac1z\)\, \delta(\eta)\, \delta\(\qthat^2-\xi_1-\xi_2-2\sqrt{\xi_1\xi_2}\cos\varphi\),
\end{align}
where we had to include also the explicit dependence on $\varphi$ as it appears in the delta function.
This result expresses the fully differential distribution in terms of the distribution
differential only in the angle $\varphi$ between $\kvec_1$ and $\kvec_2$,
and it will be used in App.~\ref{app:PairCF} to construct simplified explicit expressions for the resummed contributions.

The key object that we need is thus
\beq\label{eq:dCoffdphi}
\frac{\dd{\cal C}}{\dd\varphi}(z,\xi_1,\xi_2,Q^2,\varphi) = \theta(Q^2-4m^2)
\frac12\frac1{2\pi} \sigma_0 \int \dd\thetapair\,\delta(\thetapair)\int \dd\phi_2(q;p,\bar p)\;|{\cal M}|^2,
\eeq
where $\sigma_0=16\pi^2\as^2/Q^2$, the factor $1/2$ is the flux factor and the $1/2\pi$ comes from the $\varphi$ average.
The matrix element squared $|{\cal M}|^2$ is given in Appendix~\ref{app:ME}.
Note that because of the delta functions in Eq.~\eqref{eqP:triplediffxs2}
not all the variables of Eq.~\eqref{eq:dCoffdphi} are independent. In particular one can
write $1/z=1+\xi_1+\xi_2+2\sqrt{\xi_1\xi_2}\cos\varphi$ and use it to fix one of them in terms of the others.

We now focus on the computation of $\dd{\cal C}/\dd\varphi$. We observe that
the two-body phase space Eq.~\eqref{eq:twobody} contains two delta functions corresponding to the mass shell condition of the heavy quarks.
We write them in terms of the new variables, and get
\begin{subequations}\label{eq:onshellppbar}
\begin{align}
0=p^2-m^2 &= z_1z_2\frac{Q^2}{z}-|\pvec|^2-m^2 \\
0=\bar p^2-m^2 &= (1-z_1)(1-z_2)\frac{Q^2}{z}-|\qvec-\pvec|^2-m^2\nonumber\\
&= (1-z_1-z_2)\frac{Q^2}{z}-|\qvec|^2+2\qvec\cdot\pvec,
\end{align}
\end{subequations}
where in the last step we have used the first on-shell condition.
The second condition contains a scalar product, and thus an angle, which is not ideal as this appears in the argument of the delta function.
In order to get rid of the scalar product, we use the first condition to fix $z_2$, through the equation
\beq
z_2 = z\frac{|\pvec|^2+m^2}{z_1Q^2}
\eeq
so that the second condition becomes
\begin{align}
0=\bar p^2-m^2
&= (1-z_1)\frac{Q^2}{z}-\frac{|\pvec|^2+m^2}{z_1}-|\qvec|^2+2\qvec\cdot\pvec,
\end{align}
We can now get rid of the scalar product by introducing a new vector $\Dvec$ defined by
\beq
\pvec = z_1\qvec+\Dvec
\eeq
so that
\begin{align}
0=\bar p^2-m^2
&= (1-z_1)\frac{Q^2}{z}-\frac{|z_1\qvec+\Dvec|^2+m^2}{z_1}-|\qvec|^2+2z_1|\qvec|^2+2\qvec\cdot\Dvec\nonumber\\
&= (1-z_1)\(\frac{Q^2}{z}-|\qvec|^2\)-\frac{|\Dvec|^2+m^2}{z_1}\nonumber\\
&= (1-z_1)Q^2-\frac{|\Dvec|^2+m^2}{z_1},
\end{align}
that only depends on squared vectors (in the last step we have used $1+\qthat^2=\frac1z$).
This can be now used to fix
\beq
|\Dvec|^2 = z_1 (1-z_1)Q^2 -m^2.
\eeq
The two-body phase space can thus be rewritten as
\begin{align}\label{eq:dphi2}
  \dd\phi_2(q;p,\bar p)
  &= \frac{\dd^4p}{\(2\pi\)^3}\frac{\dd^4\bar p}{\(2\pi\)^3}\delta\(p^2-m^2\)\delta\(\bar p^2-m^2\)\(2\pi\)^4\delta^{(4)}\(q-p-\bar p\) \theta(p^0) \theta(\bar p^0) \nonumber \\
  &=\frac{\dd^4p}{4\pi^2}\delta\(p^2-m^2\) \delta\((q-p)^2-m^2\) \theta(p^0) \theta(q^0- p^0)\nonumber\\
  &=\frac{Q^2}{8\pi^2z}\delta\(z_1z_2\frac{Q^2}{z}-|\pvec|^2-m^2\) \delta\((1-z_1-z_2)\frac{Q^2}{z}-|\qvec|^2+2\qvec\cdot\pvec\) \dd z_1\, \dd z_2\, \dd^2\pvec  \nonumber\\
  &\quad\times\theta(z_1)\theta(z_2)\theta(1-z_1)\theta(1-z_2)  \nonumber\\
&=\frac1{8\pi^2} \delta\((1-z_1)Q^2-\frac{|\Dvec|^2+m^2}{z_1}\) \theta(z_1)\theta(1-z_1)\frac{\dd z_1}{z_1}\, \dd^2\Dvec  \nonumber\\
&=\frac1{16\pi^2} \theta\(z_1 (1-z_1)Q^2 -m^2\) \dd z_1\, \dd\omega \nonumber\\
&=\frac1{16\pi^2} \theta\(\sqrt{\frac14-\frac{m^2}{Q^2}}-\abs{\frac12-z_1}\) \dd z_1\, \dd\omega \nonumber\\
&=\frac1{16\pi^2} \sqrt{\frac14-\frac{m^2}{Q^2}} \sin\beta\, \dd\beta\, \dd\omega
\end{align}
where $\omega$ is the azimuthal angle of $\Dvec$ with respect to $\kvec_1+\kvec_2$.
Note that the condition $Q^2>4m^2$, needed to satisfy the theta function, is always verified in Eq.~\eqref{eq:dCoffdphi}.
If we wish to compute the $z_1$ integral numerically, it is convenient to change variable as
\beq\label{eq:z1beta}
z_1 = \frac12-\sqrt{\frac14-\frac{m^2}{Q^2}}\cos\beta,
\qquad \beta\in[0,\pi]
\eeq
which we used to obtain the last line of Eq.~\eqref{eq:dphi2}. Interestingly, in terms of these variables $|\Dvec|^2$ becomes
\beq
|\Dvec|^2 = \frac{Q^2 -4m^2}4\sin^2\beta \qquad\Rightarrow\qquad
|\Dvec| = \frac12\sqrt{Q^2 -4m^2}\sin\beta,
\eeq
where we do not need to include an absolute value, as in the allowed range $\sin\beta$ is always positive.

The form of the phase space Eq.~\eqref{eq:dphi2} is very convenient from a numerical point of view.
To be able to perform all integrations, we also need to express the matrix element squared appearing
in Eq.~\eqref{eq:dCoffdphi} in terms of the variables $\beta$ (or $z_1$) and $\omega$.
We start by rewriting
\begin{align}
z_2 &= z\frac{|z_1\qvec+\Dvec|^2+m^2}{z_1Q^2} \nonumber\\
&= z\frac{z_1^2|\qvec|^2+|\Dvec|^2+2z_1\qvec\cdot\Dvec+m^2}{z_1Q^2} \nonumber\\
&= z\[1-z_1(1-\qthat^2)+2\sqrt{\qthat^2}\sqrt{z_1 (1-z_1) -m^2/Q^2}\cos(\omega-\thetapair)\] \nonumber\\
&= z\[1-(1-\qthat^2)\(\frac12-\sqrt{\frac14-\frac{m^2}{Q^2}}\cos\beta\)+\sqrt{\qthat^2}\sqrt{1-4m^2/Q^2}\sin\beta\cos(\omega-\thetapair)\]
\end{align}
where $\thetapair=0$ for our choice of variables, Eq.~\eqref{eqP:finalstatevars}.
Finally, we will see in appendix~\ref{app:ME} that the matrix element depends on the scalar product $\kvec_2\cdot\pvec$
through the variables $T$ Eq.~\eqref{eq:T} and $U$ Eq.~\eqref{eq:U}.
We can write
\begin{align}
  \frac{\kvec_2\cdot\pvec}{Q^2}
  &=  \frac{\kvec_2\cdot(z_1\qvec+\Dvec)}{Q^2} \nonumber\\
  &=  \frac{z_1(\kvec_2^2+ \kvec_2\cdot\kvec_1)+\kvec_2\cdot\Dvec}{Q^2} \nonumber\\
  &=  z_1\xi_2+ z_1\sqrt{\xi_1\xi_2}\cos\varphi+\sqrt{\xi_2}\sqrt{\frac14-\frac{m^2}{Q^2}}\sin\beta\cos\omega',
\end{align}
where $\omega'$ is the angle between $\Dvec$ and $\kvec_2$. It is given by $\omega'=\omega+\varphi'$,
where $\varphi'$ is the angle of $\qvec=\kvec_1+\kvec_2$ with respect to $\kvec_2$, given in Eq.~\eqref{eq:varphi'}.
For the on-shell limit $\xi_2\to0$ it is also useful to write
\begin{align}
\pthat^2 &\equiv \frac{\pvec^2}{Q^2} = z_1^2\xi_1+\frac{|\Dvec|^2}{Q^2} + 2z_1 \sqrt{\xi_1}\frac{|\Dvec|}{Q}\cos\omega
\end{align}
in terms of the new phase-space variables.
In the fully on-shell limit the result simplifies further
\begin{align}
\pthat^2 &= \frac{|\Dvec|^2}{Q^2} = z_1(1-z_1)-\frac{m^2}{Q^2}.
\end{align}

\subsection{Matrix element}
\label{app:ME}

In this Appendix we report the matrix element squared for heavy quark pair production
from two off-shell gluons. This has been computed in Refs.~\cite{Catani:1990eg,Ball:2001pq}.
Here, we rewrite that result in terms of the variables that we have defined above.

The matrix element is separated into an abelian and a non-abelian parts as
\beq
\abs{\cal M}^2 = \frac1{2C_A}\abs{\cal M}^2_{\rm ab} + \frac1{4C_F}\abs{\cal M}^2_{\rm nab}
\eeq
with
\beq\label{eq:Mab}
\abs{\cal M}^2_{\rm ab} = \frac1{z^2}\[\frac1{TU} - \frac1{\xi_1\xi_2}\(1+\frac{z_2(1-z_1)}{zT}+\frac{z_1(1-z_2)}{zU}\)^2\]
\eeq
and
\begin{align}\label{eq:Mnab}
  \abs{\cal M}^2_{\rm nab}
  &= \frac1{z^2}\bigg[-\frac1{TU}+ \frac{2z}S +\frac{(T-U)(z_1-z_2)}{STU} \nonumber\\
  &\qquad\qquad+ \frac2{\xi_1\xi_2}
  \(\frac12+\frac{z_2(1-z_1)}{zT}-\frac\Delta S\)
  \(\frac12+\frac{z_1(1-z_2)}{zU}+\frac\Delta S\)
\bigg],
\end{align}
where
\beq
\Delta = \frac{z_1(1-z_2)}z - \frac{z_2(1-z_1)}z + \xi_1z_2-\xi_2z_1+\frac{z_2-z_1}{2z}+\frac{\pvec\cdot(\kvec_2-\kvec_1)}{Q^2}
\eeq
and
\begin{subequations}
\begin{align}
  S=\frac{\hat s}{Q^2} &= \frac{(k_1+k_2)^2}{Q^2} = \frac1z-\xi_1-\xi_2-2\sqrt{\xi_1\xi_2}\cos\varphi \label{eq:S}\\
  T= \frac{t-m^2}{Q^2} &= \frac{(p-k_1)^2-m^2}{Q^2} = \frac{2\kvec_1\cdot\pvec}{Q^2} - \xi_1-\frac{z_2}z\\
                       &= \frac{(\bar p-k_2)^2-m^2}{Q^2} = -\frac{2\kvec_2\cdot\pvec}{Q^2} + \xi_2+2\sqrt{\xi_1\xi_2}\cos\varphi-\frac{1-z_1}z\label{eq:T}\\
  U= \frac{u-m^2}{Q^2} &= \frac{(p-k_2)^2-m^2}{Q^2} = \frac{2\kvec_2\cdot\pvec}{Q^2} - \xi_2-\frac{z_1}z\label{eq:U}\\
                       &= \frac{(\bar p-k_1)^2-m^2}{Q^2} = -\frac{2\kvec_1\cdot\pvec}{Q^2} + \xi_1+2\sqrt{\xi_1\xi_2}\cos\varphi-\frac{1-z_2}z.
\end{align}
\end{subequations}
Note that in the case of the pair kinematics, where $q=p+\bar p$, we have $\hat s=Q^2$ and thus $S=1$.
We can use the expressions of $T$ and $U$ to rewrite
\beq
\frac{\pvec\cdot(\kvec_2-\kvec_1)}{Q^2} = \frac12\[U-T+\xi_2-\xi_1+\frac{z_1-z_2}z\]
\eeq
so that $\Delta$ simplifies to
\beq\label{eq:Deltadef}
\Delta = \frac{z_1-z_2}z + \xi_1z_2-\xi_2z_1+\frac{U+\xi_2-T-\xi_1}2.
\eeq
We also recall the relation
\beq
\hat s+t+u = 2m^2-|\kvec_1|^2-|\kvec_2|^2,
\eeq
namely
\beq
S+T+U+\xi_1+\xi_2 = 0.
\eeq
Thus, one can always express one of these variables in terms of the other four.

Note that the matrix element squared is symmetric under the simultaneous exchange
\beq
\kvec_1\leftrightarrow\kvec_2
,\qquad
z_1\leftrightarrow z_2.
\eeq
This implies that, in the pair kinematics, after integrating over the two $z_1$ and $z_2$ variables (which appear symmetrically in the phase space)
the off-shell coefficient is symmetric under the exchange of the two gluon virtualities.
Similarly, in the single-quark kinematics, the off-shell coefficient is symmetric under the exchange of the two gluon virtualities
and a sign change in the rapidity $\eta$.

\subsection{On-shell limit}

The resummation discussed in sect.~\ref{sec:channels} requires also the coefficient function with just one gluon off-shell.
This result can be obtained by simply taking the partial on-shell limit, say $\kvec_2\to0$, of the fully off-shell result.
Here we perform this limit at the level of the matrix element squared.
We will also compute the fully on-shell limit, needed for the fixed-order expansion, which also serves as a cross check.

When taking the on-shell limit, one must be careful in the choice of the parameters used to write the matrix element.
Previously, we have used the most convenient variables to obtain a compact form, but some of them are not independent of the others.
When taking an on-shell limit, any such relation must be made explicit.

We commented at the beginning of this appendix that the off-shell coefficient depends on 7 independent variables,
4 initial-state ones and 3 final-state ones.
Since the matrix element is dimensionless, we shall conveniently choose a set of dimensionless variables.
Thus, for the initial state we use the variables $z, \xi_1, \xi_2, \varphi$ defined in Eq.~\eqref{eq:initialstatevar},
while for the final state we could consider any three variables out of $\vartheta, \pthat, z_1, z_2$, namely
\beq\label{eq:fsvsets}
\vartheta, \pthat, z_1 \qquad \text{or}\qquad
\vartheta, \pthat, z_2 \qquad \text{or}\qquad
\vartheta, z_1, z_2 \qquad \text{or}\qquad
\pthat, z_1, z_2.
\eeq
The relation between these four variables is given by the equation
\beq\label{eq:pthatconstraint}
\pthat = \frac{\frac{z_1+z_2-1}z+\xi_1+\xi_2+2\sqrt{\xi_1\xi_2}\cos\varphi}{2\sqrt{\xi_1+\xi_2+2\sqrt{\xi_1\xi_2}\cos\varphi}\cos\vartheta},
\eeq
which descends from the on-shell condition $p^2=\bar p^2\; (=m^2)$.
All these choices are acceptable provided they are kept throughout the computation of the on-shell limit.
Once the on-shell limit $\kvec_2\to0$ is taken, we must also compute the average over $\varphi$ (as it is no longer well defined),
so that the remaining variables upon which the matrix element can be expressed are just 5, e.g.\ $z,\xi_1,\vartheta,z_1,z_2$
or $z,\xi_1,\pthat,z_1,z_2$.
If we also want to compute the fully on-shell limit $\kvec_{1,2}\to0$, the azimutal angle of $\pvec$ becomes arbitrary,
and the result depends on just 3 independent variables, $z,\pthat,z_1$ or $z,\pthat,z_2$
(not $z,z_1,z_2$ because they are no longer independent in the on-shell case).
Therefore, it is convenient to have $\pthat$ in our variable set, so we discard the third set of Eq.~\eqref{eq:fsvsets}.
The most convenient set is probably the last, $\pthat, z_1, z_2$, so we go for it.

When looking at the matrix element squared, Eqs.~\eqref{eq:Mab} and \eqref{eq:Mnab},
it is clear that there is a potential singularity in the on-shell limit due to the presence of a factor $1/\xi_2$.
This is harmless if the terms in rounded brackets are of order $\sqrt{\xi_2}$.
To prove this, we first expand $S,T,U$ at small $\xi_2$.
From the representations of Eqs.~\eqref{eq:S}, \eqref{eq:T} and \eqref{eq:U} we get immediately
\begin{subequations}
\begin{align}
  S &= \frac1z-\xi_1-2\sqrt{\xi_1\xi_2}\cos\varphi +\Ord(\xi_2)\\
  T &= -\frac{1-z_1}z +2\sqrt{\xi_1\xi_2}\cos\varphi -2\sqrt{\pthat^2\xi_2}\cos(\thetap-\varphi) + \Ord(\xi_2)\\
  U &= -\frac{z_1}z + 2\sqrt{\pthat^2\xi_2}\cos(\thetap-\varphi) + \Ord(\xi_2),
\end{align}
\end{subequations}
where $\thetap$ is the angle of $\pvec$ with respect to $\kvec_1$,
which coincides with $\vartheta$ in the on-shell limit $\kvec_2\to0$.

Let us start from the abelian part of the matrix element, Eq.~\eqref{eq:Mab}.
Expanding the rounded brakets at small $\xi_2$, we find
\begin{align}
1&+\frac{z_2(1-z_1)}{zT}+\frac{z_1(1-z_2)}{zU}=\nonumber\\
&= 1-\frac{z_2}{1-\frac{z}{1-z_1}\(2\sqrt{\xi_1\xi_2}\cos\varphi -2\sqrt{\xi_2\pthat^2}\cos(\thetap-\varphi)\)}
  -\frac{1-z_2}{1-\frac{z}{z_1}2\sqrt{\xi_2\pthat^2}\cos(\thetap-\varphi)}  + \Ord(\xi_2)\nonumber\\
&= -2z\sqrt{\xi_2} \[\frac{z_2}{1-z_1}\sqrt{\xi_1}\cos\varphi +\(\frac{1-z_2}{z_1}-\frac{z_2}{1-z_1}\)\sqrt{\pthat^2}\cos(\thetap-\varphi)\] +\Ord(\xi_2),
\end{align}
which is indeed of order $\sqrt{\xi_2}$.
Averaging over $\varphi$, the abelian part of the matrix element squared becomes
\begin{align}
\abs{\cal M}^2_{\rm ab} &\overset{\kvec_2\to0}= \int_0^{2\pi}\frac{\dd\varphi}{2\pi}\,
\frac1{z^2}\[\frac1{TU} - \frac{4z^2}{\xi_1}\(\frac{z_2}{1-z_1}\sqrt{\xi_1}\cos\varphi +\(\frac{1-z_2}{z_1}-\frac{z_2}{1-z_1}\)\sqrt{\pthat^2}\cos(\thetap-\varphi)\)^2\]\nonumber\\
&= \frac1{z_1(1-z_1)} - \frac{2}{\xi_1}\bigg[
\(\frac{z_2}{1-z_1}\)^2\xi_1 +\(\frac{1-z_2}{z_1}-\frac{z_2}{1-z_1}\)^2\pthat^2\nonumber\\
&\qquad\qquad\qquad\qquad + \frac{z_2}{1-z_1}\(\frac{1-z_2}{z_1}-\frac{z_2}{1-z_1}\) 2\sqrt{\xi_1\pthat^2}\cos\vartheta\bigg],
\end{align}
where we have replaced $\thetap$ with $\vartheta$ as they now coincide.
Note that we are using more variables than needed. Indeed, using Eq.~\eqref{eq:pthatconstraint}
we can rewrite $\cos\vartheta$ in terms of other variables. In particular, in the $\xi_2\to0$ limit
it is easy to obtain from Eq.~\eqref{eq:pthatconstraint} the relation
\beq\label{eq:pthatconstraint2}
2\sqrt{\xi_1\pthat^2}\cos\vartheta = \frac{z_1+z_2-1}z+\xi_1,
\eeq
from which we finally find
\begin{align}
\abs{\cal M}^2_{\rm ab} &\overset{\kvec_2\to0}= 
\frac{1-2 z_2(1-z_2)}{z_1(1-z_1)} - \frac{2}{\xi_1}
\(\frac{1-z_1-z_2}{z_1(1-z_1)}\)^2\[\pthat^2 - \frac{z_1z_2}z\].
\end{align}
For the non-abelian part, Eq.~\eqref{eq:Mnab},
let us start by expanding $\Delta$, Eq.~\eqref{eq:Deltadef}, to order $\sqrt{\xi_2}$:
\begin{align}
\Delta
&= \(\xi_1-\frac1z\)\(z_2-\frac12\)
+\sqrt{\xi_2}\(2\sqrt{\pthat^2}\cos(\thetap-\varphi) -\sqrt{\xi_1}\cos\varphi \) +\Ord(\xi_2) .
\end{align}
The rounded brakets of Eq.~\eqref{eq:Mnab} become
\begin{align}
\frac12&+\frac{z_2(1-z_1)}{zT}-\frac\Delta S= \nonumber\\
  &= 2z\sqrt{\xi_2}\[\(\frac{z_2}{1-z\xi_1}-\frac{z_2}{1-z_1}\)\sqrt{\xi_1}\cos\varphi
    -\(\frac1{1-z\xi_1}-\frac{z_2}{1-z_1}\)\sqrt{\pthat^2}\cos(\thetap-\varphi) \]
+\Ord(\xi_2) \nonumber\\
\frac12&+\frac{z_1(1-z_2)}{zU} + \frac\Delta S= \nonumber\\
  &= 2z\sqrt{\xi_2}\[-\frac{z_2}{1-z\xi_1}\sqrt{\xi_1}\cos\varphi
    +\(\frac1{1-z\xi_1}-\frac{1-z_2}{z_1}\)\sqrt{\pthat^2}\cos(\thetap-\varphi) \]
+\Ord(\xi_2) ,
\end{align}
so that we can finally find the partial on-shell limit of the non-abelian part of the matrix element,
\begin{align}
  \abs{\cal M}^2_{\rm nab}
  &\overset{\kvec_2\to0}= \int_0^{2\pi}\frac{\dd\varphi}{2\pi}\,
    \bigg\{-\frac1{z^2TU}+ \frac{2}{zS} +\frac{(T-U)(z_1-z_2)}{z^2STU} \nonumber\\
  &\qquad\qquad\qquad+ \frac8{\xi_1}
    \[\(\frac{z_2}{1-z\xi_1}-\frac{z_2}{1-z_1}\)\sqrt{\xi_1}\cos\varphi
    -\(\frac1{1-z\xi_1}-\frac{z_2}{1-z_1}\)\sqrt{\pthat^2}\cos(\thetap-\varphi) \]\nonumber\\
  &\qquad\qquad\qquad\quad\times \[-\frac{z_2}{1-z\xi_1}\sqrt{\xi_1}\cos\varphi
    +\(\frac1{1-z\xi_1}-\frac{1-z_2}{z_1}\)\sqrt{\pthat^2}\cos(\thetap-\varphi) \]\bigg\} \nonumber\\
  &=-\frac1{z_1(1-z_1)} + \frac2{1-z\xi_1} + \frac{(1-2z_1)(z_2-z_1)}{z_1(1-z_1)(1-z\xi_1)}\nonumber\\
  &\quad+ \frac4{\xi_1(1-z\xi_1)^2}
    \bigg[ \frac{(z_1-z\xi_1) z_2^2}{1-z_1} \xi_1
    -\(1-\frac{z_2(1-z\xi_1)}{1-z_1}\) \(1-\frac{(1-z_2)(1-z\xi_1)}{z_1}\) \pthat^2
    \nonumber\\
  &\qquad\qquad + \left\{\(1-\frac{1-z\xi_1}{1-z_1}\) \(1-\frac{(1-z_2)(1-z\xi_1)}{z_1}\)+1-\frac{z_2(1-z\xi_1)}{1-z_1}\right\}z_2\sqrt{\xi_1\pthat^2}\cos\vartheta
    \bigg].
\end{align}
Using again Eq.~\eqref{eq:pthatconstraint2} we finally get
\begin{align}
  \abs{\cal M}^2_{\rm nab}
  &=-\frac1{z_1(1-z_1)} + \frac2{1-z\xi_1} + \frac{(1-2z_1)(z_2-z_1)}{z_1(1-z_1)(1-z\xi_1)}\nonumber\\
  &\quad+ \frac{2z_2}{(1-z\xi_1)^2}
    \bigg[ 2\frac{(z_1-z\xi_1) z_2}{1-z_1}
    + \(1-\frac{1-z\xi_1}{1-z_1}\) \(1-\frac{(1-z_2)(1-z\xi_1)}{z_1}\)+1-\frac{z_2(1-z\xi_1)}{1-z_1}\bigg]
    \nonumber\\
  &\quad+ \frac2{\xi_1(1-z\xi_1)^2}
    \bigg[ -2\(1-\frac{z_2(1-z\xi_1)}{1-z_1}\) \(1-\frac{(1-z_2)(1-z\xi_1)}{z_1}\) \pthat^2
    \nonumber\\
  &\qquad\qquad + \left\{\(1-\frac{1-z\xi_1}{1-z_1}\) \(1-\frac{(1-z_2)(1-z\xi_1)}{z_1}\)+1-\frac{z_2(1-z\xi_1)}{1-z_1}\right\}z_2\frac{z_1+z_2-1}z
    \bigg].
\end{align}

We can now further take the limit $\kvec_1\to0$ to obtain the fully on-shell result.
This is useful as a cross-check as it must coincide with the on-shell computation, see e.g.~\cite{ParticleDataGroup:2022pth}.
In the on-shell limit, $z$, $z_1$ and $z_2$ are no longer independent, as one can see from Eq.~\eqref{eq:pthatconstraint2}.
Moreover, the angle $\vartheta$ becomes arbitrary (the reference vector $\kvec_1$ does not exist anymore),
so an average over $\vartheta$ must be taken.
From Eq.~\eqref{eq:pthatconstraint2} we can write
\beq
z_2 = 1-z_1+2z\sqrt{\pthat^2\xi_1}\cos\vartheta+\Ord(\xi_1)
\eeq
from which we find
\begin{align}
\abs{\cal M}^2_{\rm ab}
  &\overset{\kvec_{1,2}\to0}= \frac1{z_1(1-z_1)}\[1+4z\pthat^2\(1-\frac{z\pthat^2}{z_1(1-z_1)}\)\] - 2 \\
\abs{\cal M}^2_{\rm nab}
  &\overset{\kvec_{1,2}\to0}= 4z_1(1-z_1) -2 - 8z\pthat^2 \( 1-\frac{z\pthat^2}{z_1(1-z_1)} \),
\end{align}
where we can also write $\pthat^2=\frac{z_1(1-z_1)}z-\frac{m^2}{Q^2}$.
We have verified that this result is in agreement with on-shell computations~\cite{ParticleDataGroup:2022pth}.
For completeness, we also report the on-shell matrix element for the $q\bar q$ channel:
\begin{align}
\abs{\cal M}^2_{\rm q\bar q}
  &= \frac{C_F}{C_A}\(1-2z\pthat^2\) .
\end{align}

\section{Simplifications in the resummation formulae for pair kinematics}
\label{app:PairCF}

When considering pair kinematics we fix $q\equiv p+\bar p$.
Because of momentum conservation, we also have $q=k_1+k_2$.
Therefore, each component of $q$ is fixed in terms of the initial state variables.
It follows that the off-shell coefficient function factorizes as in Eq.~\eqref{eqP:triplediffxs2},
that we report here for convenience
\begin{align}\label{eq:4diff}
\frac{Q^4\,\dd{\cal C}}{\dd Q^2\,\dd \eta\, \dd \qt^2\, \dd\varphi}(z,\xi_1,\xi_2,Q^2, \eta,\qt^2,\varphi)
  &=  \frac{\dd{\cal C}}{\dd\varphi}(z,\xi_1,\xi_2,Q^2,\varphi) \nonumber\\
  &\times \delta\(1+\qthat^2-\frac1z\) \delta(\eta) \delta\(\qthat^2-\xi_1-\xi_2-2\sqrt{\xi_1\xi_2}\cos\varphi\),
\end{align}
where the differential coefficient is given in terms of a more integrated one times three delta functions.
These delta functions can be used to compute the integrations in the resummation formulae
of sect.~\ref{sec:channels}.
In this appendix, we exploit this to present simplified resummed expressions, that we have implemented in the numerical code \hell.

From Eq.~\eqref{eq:4diff} we can obtain immediately the triple differential off-shell coefficient function
by integrating in $\varphi$ using the last delta function
\begin{align}\label{eq:3diff}
\frac{Q^4\,\dd{\cal C}}{\dd Q^2\,\dd \eta\, \dd \qt^2}(z,\xi_1,\xi_2,Q^2, \eta,\qt^2)
  &= 
\frac{\theta\(1-\abs{\frac{\qthat^2-\xi_1-\xi_2}{2\sqrt{\xi_1\xi_2}}}\)}{\sqrt{4\xi_1\xi_2-(\qthat^2-\xi_1-\xi_2)^2}}
\delta\(1+\qthat^2-\frac1z\) \delta(\eta)
 \nonumber\\
  &\times 
\[\frac{\dd{\cal C}}{\dd\varphi}\(z,\xi_1,\xi_2,Q^2,\bar\varphi\)
+\frac{\dd{\cal C}}{\dd\varphi}\(z,\xi_1,\xi_2,Q^2,2\pi-\bar\varphi\)\] \\
\bar\varphi&=\cos^{-1}\frac{\qthat^2-\xi_1-\xi_2}{2\sqrt{\xi_1\xi_2}}, \qquad 0\leq\bar\varphi\leq\pi,
\end{align}
where the $\varphi$-differential distribution is evaluated at specific values of $\varphi$.
Note however that integrating over $\varphi$ immediately is not always the best strategy.
For instance, when we take the partial on-shell limit $\xi_2\to0$
we obtain
\begin{align}\label{eq:4diff2}
\frac{Q^4\,\dd{\cal C}}{\dd Q^2\,\dd \eta\, \dd \qt^2\, \dd\varphi}(z,\xi_1,0,Q^2, \eta,\qt^2,\varphi)
  &=  \frac{\dd{\cal C}}{\dd\varphi}(z,\xi_1,0,Q^2,\varphi)
  \delta\(1+\qthat^2-\frac1z\) \delta(\eta) \delta\(\qthat^2-\xi_1\),
\end{align}
where the delta functions do not depend on $\varphi$ anymore, thus making its integration trivial
\begin{align}\label{eq:3diff3}
\frac{Q^4\,\dd{\cal C}}{\dd Q^2\,\dd \eta\, \dd \qt^2}(z,\xi_1,0,Q^2, \eta,\qt^2)
  &=  {\cal C}(z,\xi_1,0,Q^2)
  \delta\(1+\qthat^2-\frac1z\) \delta(\eta) \delta\(\qthat^2-\xi_1\).
\end{align}
This result is needed for the auxiliary function Eq.~\eqref{eq:resCaux},
and also for the subtraction of the plus distributions in the perturbative expansion of the resummed result, see sect.~\ref{sec:matching}.

We can now use these results in the resummation formulae, using the delta functions to perform integrations explicitly when possible.
As far as the auxiliary function Eq.~\eqref{eq:resCaux} is concerned, we can start from Eq.~\eqref{eq:3diff3}
and use the $\delta\(\qthat^2-\xi_1\)$ to compute the $\xi_1$ integration. The result is
\begin{align}\label{eq:resCaux1}
  \frac{Q^4\,\dd C_{\rm aux\,\pm}}{\dd Q^2\dd y\dd\qt^2}\(x,Q^2,y,\qt^2,\as, \frac{Q^2}{\muf^2}\)
  &= \frac1{1+\qthat^2} \,
  {\cal C}\(\frac1{1+\qthat^2},\qthat^2,0,Q^2\)
  U'_{\rm reg}\(x(1+\qthat^2),\qt^2,\muf^2\) \nonumber\\
  &\quad\times\delta\(y\pm\frac12\log\frac1{x(1+\qthat^2)}\),
\end{align}
which does not contain any further integration. Note the presence of a delta function in the result,
which can be used in the cross section to integrate over parton distributions.
The fixed-order expansion of Eq.~\eqref{eq:resCaux1} is given according to Eq.~\eqref{eq:resCauxFO} by
\begin{align}\label{eq:resCaux1FO}
  \frac{Q^4\,\dd C_{\rm aux\,\pm}}{\dd Q^2\dd y\dd\qt^2}\(x,Q^2,y,\qt^2,\as, \frac{Q^2}{\muf^2}\)
  &= \frac1{1+\qthat^2} \,
  {\cal C}\(\frac1{1+\qthat^2},\qthat^2,0,Q^2\)\delta\(y\pm\frac12\log\frac1{x(1+\qthat^2)}\)
    \nonumber\\
  &\times \bigg\{\as(\muf^2)P_0\(x(1+\qthat^2)\)\(\frac1{\qthat^2}\)_+ \nonumber\\
  &\qquad+ \as^2(\muf^2)\Bigg[P_1\(x(1+\qthat^2)\) \(\frac1{\qthat^2}\)_+ \nonumber\\
  &\qquad\qquad\quad+ \Big(P_{00}\(x(1+\qthat^2)\)-\beta_0P_0\(x(1+\qthat^2)\)\Big) \(\frac{\log\frac{\qt^2}{\muf^2}}{\qthat^2}\)_+\Bigg]\nonumber\\
  &\qquad+ \Ord(\as^3)\bigg\}.
\end{align}
We observe that, for this auxiliary function, the expansion is a distribution in $\qt^2$.
This is not an issue: the triple differential distribution is interesting only
for non-zero values of $\qt^2$, and indeed any measurement will require a $\qt^2$ grater than some resolution cutoff.
Should one be interested in integrating over $\qt^2$ down to zero, either to compute the integrated distribution or to obtain
a binned version of the $\qt^2$ distribution, one has simply to take care of using the plus distribution
in the integration.

We now move to the regular function Eq.~\eqref{eq:resCreg}.
We get
\begin{align}
  \frac{Q^4\,\dd C_{\rm reg}}{\dd Q^2\dd y\dd\qt^2}\(x,Q^2, y,\qt^2,\as, \frac{Q^2}{\muf^2}\)
  &= \frac1{1+\qthat^2}
    \int_0^\infty \dd\xi_1 \int_0^\infty \dd\xi_2 \int_0^{2\pi}\dd \varphi \,  \frac{\dd{\cal C}}{\dd\varphi}\(\frac1{1+\qthat^2},\xi_1,\xi_2,Q^2,\varphi\)\nonumber\\
  &\times U'_{\rm reg}\(\sqrt{x(1+\qthat^2)}e^{y},Q^2\xi_1,\muf^2\)\, U'_{\rm reg}\(\sqrt{x(1+\qthat^2)}e^{-y},Q^2\xi_2,\muf^2\) \nonumber\\
  &\times \delta\(\qthat^2-\xi_1-\xi_2-2\sqrt{\xi_1\xi_2}\cos\varphi\) \,\theta\(\frac{e^{-2|y|}}x-1-\qthat^2\) \label{eq:resCreg1}\\
  &= \frac{\theta\(\frac{e^{-2|y|}}x-1-\qthat^2\)}{1+\qthat^2}
    \int_0^\infty \dd\xi_1 \int_0^\infty \dd\xi_2\, \frac{\theta\(1-\abs{\frac{\qthat^2-\xi_1-\xi_2}{2\sqrt{\xi_1\xi_2}}}\)}{\sqrt{4\xi_1\xi_2-(\qthat^2-\xi_1-\xi_2)^2}} \nonumber\\
  &\times U'_{\rm reg}\(\sqrt{x(1+\qthat^2)}e^{y},Q^2\xi_1,\muf^2\)\, U'_{\rm reg}\(\sqrt{x(1+\qthat^2)}e^{-y},Q^2\xi_2,\muf^2\) \nonumber\\
  &\times 
\[\frac{\dd{\cal C}}{\dd\varphi}\(\frac1{1+\qthat^2},\xi_1,\xi_2,Q^2,\bar\varphi\)
+\frac{\dd{\cal C}}{\dd\varphi}\(\frac1{1+\qthat^2},\xi_1,\xi_2,Q^2,2\pi-\bar\varphi\)\], \label{eq:resCreg2}
\end{align}
where we have used Eq.~\eqref{eq:4diff} in the first step and Eq.~\eqref{eq:3diff} in the second step.
Note that the theta function inside the integration can be recast in the (physically obvious) constraint
\beq\label{eq:xi12constraint}
\(\sqrt{\xi_1}-\sqrt{\xi_2}\)^2\leq \qthat^2\leq \(\sqrt{\xi_1}+\sqrt{\xi_2}\)^2.
\eeq
The integration over $\xi_{1,2}$ shall be performed numerically.
The fixed-order expansion of Eq.~\eqref{eq:resCreg2} can be computed according to Eq.~\eqref{eq:resCregFO}.
To this end, it is better to start from Eq.~\eqref{eq:resCreg1}, to obtain
\begin{align}\label{eq:CregTripleExpanded}
  \frac{Q^4\,\dd C_{\rm reg}}{\dd Q^2\dd y\dd\qt^2}&\(x,Q^2,y,\qt^2,\as, \frac{Q^2}{\muf^2}\) \nonumber\\
  &= \frac1{1+\qthat^2}
    \int_0^\infty \dd\xi_1 \int_0^\infty \dd\xi_2 \int_0^{2\pi}\dd\varphi \,  \frac{\dd{\cal C}}{\dd\varphi}\(\frac1{1+\qthat^2},\xi_1,\xi_2,Q^2,\varphi\)\nonumber\\
  &\times \[\as^2(\muf^2)\plus{\frac1{\xi_1}}\plus{\frac1{\xi_2}}P_0\(\sqrt{x(1+\qthat^2)}e^{y}\)\, P_0\(\sqrt{x(1+\qthat^2)}e^{-y}\)+\Ord(\as^3)\] \nonumber\\
  &\times \delta\(\qthat^2-\xi_1-\xi_2-2\sqrt{\xi_1\xi_2}\cos\varphi\) \,\theta\(\frac{e^{-2|y|}}x-1-\qthat^2\)\\
  &= \frac1{1+\qthat^2}\theta\(\frac{e^{-2|y|}}x-1-\qthat^2\)\nonumber\\
  &\times \[\as^2(\muf^2)P_0\(\sqrt{x(1+\qthat^2)}e^{y}\)\, P_0\(\sqrt{x(1+\qthat^2)}e^{-y}\)+\Ord(\as^3)\]
  \int_0^{2\pi}\dd\varphi\, I
\end{align}
having defined
\begin{align}\label{eq:I}
 I &=\int_0^\infty \frac{\dd\xi_1}{\xi_1} \int_0^\infty \frac{\dd\xi_2}{\xi_2}
    \Bigg[  \frac{\dd{\cal C}}{\dd\varphi}\(\frac1{1+\qthat^2},\xi_1,\xi_2,Q^2,\varphi\)
    \delta\(\qthat^2-\xi_1-\xi_2-2\sqrt{\xi_1\xi_2}\cos\varphi\) \nonumber\\
  &\qquad\qquad\qquad\qquad\qquad\qquad - \frac{\dd{\cal C}}{\dd\varphi}\(\frac1{1+\qthat^2},\xi_1,0,Q^2,\varphi\)
    \delta\(\qthat^2-\xi_1\) \theta(1-\xi_2) \nonumber\\
  &\qquad\qquad\qquad\qquad\qquad\qquad - \frac{\dd{\cal C}}{\dd\varphi}\(\frac1{1+\qthat^2},0,\xi_2,Q^2,\varphi\)
    \delta\(\qthat^2-\xi_2\) \theta(1-\xi_1) \nonumber\\
  &\qquad\qquad\qquad\qquad\qquad\qquad + \frac{\dd{\cal C}}{\dd\varphi}\(\frac1{1+\qthat^2},0,0,Q^2,\varphi\)
    \delta\(\qthat^2\) \theta(1-\xi_1) \theta(1-\xi_2)\Bigg]
\end{align}
This result is not immediately usable as delta functions still appear explicitly, and the cancellation
of singularities in $\xi_1,\xi_2=0$ requires integrating over these delta functions in a proper order.

To do so, we make some observations.
First, when $\xi_1$ or $\xi_2$ is zero, the coefficient does no longer depend on $\varphi$ in principle.
However, when considering the limit $\xi_{1,2}\to0$, a dependence on $\varphi$ remains. So, for later convenience, we keep the last argument
having in mind a limit procedure.
Second, the last term is proportional to $\delta(\qthat^2)$, which is zero everywhere in the $\qthat^2$ distribution,
except for a single point. This point is interesting only for computing the cumulative distribution between $\qthat^2=0$
and some given value, but in this case it is more convenient to consider the integrated distribution
and subtract from it the integral from that value to infinity.
Therefore, for our purposes, we can assume $\qthat^2>0$ and ignore the last line.
Finally, we observe that the integrand is symmetric for the exchange $\xi_1\leftrightarrow\xi_2$, as a consequence of
the analogous symmetry of the function $\dd{\cal C}/\dd\varphi$.

We can separate the integration region into 4 subregions, divided by the lines $\xi_1=1$ and $\xi_2=1$.
As a result we can write
\begin{align}
I &= I_1 + I_2 + I_3 + I_4 \\
I_1 &= \int_0^1 \frac{\dd\xi_1}{\xi_1} \int_0^1 \frac{\dd\xi_2}{\xi_2}
    \Bigg[\frac{\dd{\cal C}}{\dd\varphi}\(\frac1{1+\qthat^2},\xi_1,\xi_2,Q^2,\varphi\)
    \delta\(\qthat^2-\xi_1-\xi_2-2\sqrt{\xi_1\xi_2}\cos\varphi\) \nonumber\\
  &\qquad\qquad\qquad\qquad\quad - \frac{\dd{\cal C}}{\dd\varphi}\(\frac1{1+\qthat^2},\xi_1,0,Q^2\) \delta\(\qthat^2-\xi_1\) \nonumber\\
  &\qquad\qquad\qquad\qquad\quad - \frac{\dd{\cal C}}{\dd\varphi}\(\frac1{1+\qthat^2},0,\xi_2,Q^2\) \delta\(\qthat^2-\xi_2\) \Bigg] \\
I_2 &= \int_0^1 \frac{\dd\xi_1}{\xi_1} \int_1^\infty \frac{\dd\xi_2}{\xi_2}
    \Bigg[\frac{\dd{\cal C}}{\dd\varphi}\(\frac1{1+\qthat^2},\xi_1,\xi_2,Q^2,\varphi\)
    \delta\(\qthat^2-\xi_1-\xi_2-2\sqrt{\xi_1\xi_2}\cos\varphi\) \nonumber\\
  &\qquad\qquad\qquad\qquad\quad - \frac{\dd{\cal C}}{\dd\varphi}\(\frac1{1+\qthat^2},0,\xi_2,Q^2\) \delta\(\qthat^2-\xi_2\) \Bigg] \\
I_3 &= \int_1^\infty \frac{\dd\xi_1}{\xi_1} \int_0^1 \frac{\dd\xi_2}{\xi_2}
    \Bigg[\frac{\dd{\cal C}}{\dd\varphi}\(\frac1{1+\qthat^2},\xi_1,\xi_2,Q^2,\varphi\)
    \delta\(\qthat^2-\xi_1-\xi_2-2\sqrt{\xi_1\xi_2}\cos\varphi\) \nonumber\\
  &\qquad\qquad\qquad\qquad\quad - \frac{\dd{\cal C}}{\dd\varphi}\(\frac1{1+\qthat^2},\xi_1,0,Q^2\) \delta\(\qthat^2-\xi_1\) \Bigg] \\
I_4 &= \int_1^\infty \frac{\dd\xi_1}{\xi_1} \int_1^\infty \frac{\dd\xi_2}{\xi_2}\,
    \frac{\dd{\cal C}}{\dd\varphi}\(\frac1{1+\qthat^2},\xi_1,\xi_2,Q^2,\varphi\)
    \delta\(\qthat^2-\xi_1-\xi_2-2\sqrt{\xi_1\xi_2}\cos\varphi\) .
\end{align}
The $\xi_1\leftrightarrow\xi_2$ symmetry implies $I_2=I_3$, and further allows us to write $I_1$
\begin{align}\label{eq:I1v2}
I_1 &= 2\int_0^1 \frac{\dd\xi_1}{\xi_1} \int_{\xi_1}^1 \frac{\dd\xi_2}{\xi_2}
    \Bigg[\frac{\dd{\cal C}}{\dd\varphi}\(\frac1{1+\qthat^2},\xi_1,\xi_2,Q^2,\varphi\)
    \delta\(\qthat^2-\xi_1-\xi_2-2\sqrt{\xi_1\xi_2}\cos\varphi\) \nonumber\\
  &\qquad\qquad\qquad\qquad\quad - \frac{\dd{\cal C}}{\dd\varphi}\(\frac1{1+\qthat^2},\xi_1,0,Q^2\) \delta\(\qthat^2-\xi_1\) \nonumber\\
  &\qquad\qquad\qquad\qquad\quad - \frac{\dd{\cal C}}{\dd\varphi}\(\frac1{1+\qthat^2},0,\xi_2,Q^2\) \delta\(\qthat^2-\xi_2\) \Bigg].
\end{align}
as an integral over a triangle. In this way, only one subtraction is needed to make the integral finite; the other one is a finite integrable contribution.

As these integrals have to be further integrated in $\varphi$, one would be tempted to perform this integration first, before proceeding
to $\xi_{1,2}$ integration.
This seems advantageous because the ``full'' delta function can be easily solved for $\varphi$ (this is what we have already done before)
and the subtraction terms are $\varphi$-independent and thus the integral is trivial.
However, proceeding in this way may potentially lead to numerical instabilities.
Consider for instance $I_2$, where the subtraction is needed to regulate the $\xi_1$ integral in $\xi_1=0$.
After integrating analytically over $\varphi$, the remaining $\xi_2$ integration shall be done analytically (using the delta function)
for the subtraction term, but numerically (as we have already used the delta function) for the first term.
The cancellation between the two terms is then realised after a numerical integration, which may be dangerous.
To make the cancellation smoother, it is much safer to use the delta functions to fix the same variable ($\xi_2$)
in the first term and in the subtraction term. The same holds for the $I_1$ integral, in the representation Eq.~\eqref{eq:I1v2}.

To do so, we need to solve the delta function for $\xi_2$. The zeros of the argument are given by
\beq
\sqrt{\xi_2^\pm} = -\sqrt{\xi_1}\cos\varphi \pm \sqrt{\qthat^2-\xi_1(1-\cos^2\varphi)}.
\eeq
It is then convenient to change integration variable to $\sqrt{\xi_2}$. We get, for a generic function $F(\xi_2)$,
\begin{align}
  \int \dd\xi_2\,F(\xi_2)\, \delta\(\qthat^2-\xi_1-\xi_2-2\sqrt{\xi_1\xi_2}\cos\varphi\) 
  =\frac{\theta\(\qthat^2-\xi_1(1-\cos^2\varphi)\)}{\sqrt{\qthat^2-\xi_1(1-\cos^2\varphi)}}\[\sqrt{\xi_2^+}F(\xi_2^+) + \sqrt{\xi_2^-}F(\xi_2^-)\]
\end{align}
where the denominator comes from the derivative of the argument of the delta function.
According to this result we can find
\begin{align}
I_1 &= 2 \int_0^1 \frac{\dd\xi_1}{\xi_1} 
    \Bigg[\frac{\theta\(\qthat^2-\xi_1(1-\cos^2\varphi)\)}{\sqrt{\qthat^2-\xi_1(1-\cos^2\varphi)}}
      \Bigg(\frac1{\sqrt{\xi_2^+}}\frac{\dd{\cal C}}{\dd\varphi}\(\frac1{1+\qthat^2},\xi_1,\xi_2^+,Q^2,\varphi\) \theta\(1-\sqrt{\xi_2^+}\) \theta\(\sqrt{\xi_2^+}-\sqrt{\xi_1}\) \nonumber\\
    &\hspace{15em}
    + \frac1{\sqrt{\xi_2^-}}\frac{\dd{\cal C}}{\dd\varphi}\(\frac1{1+\qthat^2},\xi_1,\xi_2^-,Q^2,\varphi\) \theta\(1-\sqrt{\xi_2^-}\) \theta\(\sqrt{\xi_2^-}-\sqrt{\xi_1}\)
      \Bigg) \nonumber\\
  &\qquad\qquad\qquad\quad - \frac1{\qthat^2}\frac{\dd{\cal C}}{\dd\varphi}\(\frac1{1+\qthat^2},0,\qthat^2,Q^2\)  \theta\(1-\qthat^2\)  \theta\(\qthat^2-\xi_1\) \Bigg]\nonumber\\
&\quad + \frac2{\qthat^2}\frac{\dd{\cal C}}{\dd\varphi}\(\frac1{1+\qthat^2},\qthat^2,0,Q^2\) \log\qthat^2 \theta\(1-\qthat^2\)\\
I_2 &= \int_0^1 \frac{\dd\xi_1}{\xi_1} 
    \Bigg[\frac{\theta\(\qthat^2-\xi_1(1-\cos^2\varphi)\)}{\sqrt{\qthat^2-\xi_1(1-\cos^2\varphi)}}
      \Bigg(\frac1{\sqrt{\xi_2^+}}\frac{\dd{\cal C}}{\dd\varphi}\(\frac1{1+\qthat^2},\xi_1,\xi_2^+,Q^2,\varphi\) \theta\(\sqrt{\xi_2^+}-1\) \nonumber\\
    &\hspace{15em}
    + \frac1{\sqrt{\xi_2^-}}\frac{\dd{\cal C}}{\dd\varphi}\(\frac1{1+\qthat^2},\xi_1,\xi_2^-,Q^2,\varphi\) \theta\(\sqrt{\xi_2^-}-1\)
      \Bigg) \nonumber\\
  &\qquad\qquad\qquad - \frac1{\qthat^2}\frac{\dd{\cal C}}{\dd\varphi}\(\frac1{1+\qthat^2},0,\qthat^2,Q^2\) \theta\(\qthat^2-1\) \Bigg].
\end{align}
It is easy to check that the explicit integrals in $\xi_1$ are finite as $\xi_1\to0$. Indeed in this limit $\sqrt{\xi_2^\pm}\to\pm\sqrt{\qthat^2}$,
so the $\xi_2^-$ contributions die due to the theta functions, and the $\xi_2^+$ contributions become identical to the subtraction terms,
thus making the square bracket vanishing in the limit.

The last integral, $I_4$, can be performed similarly to $I_1$. However, because here there are no subtraction terms,
it is possible to use the delta function for any variable, and in this case it may be convenient to do it for $\varphi$.

Note that $I$ can be simplified as
\begin{align}
I &= I_1+2I_2+I_4 \nonumber\\
  &= 2 \int_0^1 \frac{\dd\xi_1}{\xi_1} 
    \Bigg[\frac{\theta\(\qthat^2-\xi_1(1-\cos^2\varphi)\)}{\sqrt{\qthat^2-\xi_1(1-\cos^2\varphi)}}
      \Bigg(\frac1{\sqrt{\xi_2^+}}\frac{\dd{\cal C}}{\dd\varphi}\(\frac1{1+\qthat^2},\xi_1,\xi_2^+,Q^2,\varphi\) \theta\(\sqrt{\xi_2^+}-\sqrt{\xi_1}\) \nonumber\\
    &\hspace{15em}
    + \frac1{\sqrt{\xi_2^-}}\frac{\dd{\cal C}}{\dd\varphi}\(\frac1{1+\qthat^2},\xi_1,\xi_2^-,Q^2,\varphi\) \theta\(\sqrt{\xi_2^-}-\sqrt{\xi_1}\)
      \Bigg) \nonumber\\
  &\qquad\qquad\qquad\quad - \frac1{\qthat^2}\frac{\dd{\cal C}}{\dd\varphi}\(\frac1{1+\qthat^2},0,\qthat^2,Q^2\)  \theta\(\qthat^2-\xi_1\) \Bigg]\nonumber\\
&\quad + \frac2{\qthat^2}\frac{\dd{\cal C}}{\dd\varphi}\(\frac1{1+\qthat^2},\qthat^2,0,Q^2\) \log\qthat^2 \theta\(1-\qthat^2\) + I_4.
\end{align}
Note also that the inequality $\sqrt{\xi_2^-}>\sqrt{\xi_1}$ required by the theta function in the second line implies
\beq
-\cos\varphi > 1+\sqrt{\frac{\qthat^2}{\xi_1}-1+\cos^2\varphi}
\eeq
which is clearly impossible, so the result simplifies further
\begin{align}
I &= 2 \int_0^1 \frac{\dd\xi_1}{\xi_1} 
    \Bigg[\frac{\theta\(\qthat^2-\xi_1(1-\cos^2\varphi)\)}{\sqrt{\qthat^2-\xi_1(1-\cos^2\varphi)}}
    \frac1{\sqrt{\xi_2^+}}\frac{\dd{\cal C}}{\dd\varphi}\(\frac1{1+\qthat^2},\xi_1,\xi_2^+,Q^2,\varphi\) \theta\(\frac{\qthat^2}{2(1+\cos\varphi)}-\xi_1\) \nonumber\\
  &\qquad\qquad\qquad\quad - \frac1{\qthat^2}\frac{\dd{\cal C}}{\dd\varphi}\(\frac1{1+\qthat^2},0,\qthat^2,Q^2\)  \theta\(\qthat^2-\xi_1\) \Bigg]\nonumber\\
&\quad + \frac2{\qthat^2}\frac{\dd{\cal C}}{\dd\varphi}\(\frac1{1+\qthat^2},\qthat^2,0,Q^2\) \log\qthat^2 \theta\(1-\qthat^2\) + I_4,
\end{align}
where we have also traded the $\sqrt{\xi_2^+}>\sqrt{\xi_1}$ condition for a simpler condition on $\xi_1$.
The second theta function is more stringent than the first one, so the first one can be dropped.
The result above can thus be written as
\begin{align}
I &= 2 \int_0^{\frac{\qthat^2}{2(1+\cos\varphi)}} \frac{\dd\xi_1}{\xi_1} 
    \Bigg[\frac1{\sqrt{\qthat^2-\xi_1(1-\cos^2\varphi)}}
    \frac1{\sqrt{\xi_2^+}}\frac{\dd{\cal C}}{\dd\varphi}\(\frac1{1+\qthat^2},\xi_1,\xi_2^+,Q^2,\varphi\) - \frac1{\qthat^2}\frac{\dd{\cal C}}{\dd\varphi}\(\frac1{1+\qthat^2},0,\qthat^2,Q^2\) \Bigg]\nonumber\\
&\quad + \frac2{\qthat^2}\frac{\dd{\cal C}}{\dd\varphi}\(\frac1{1+\qthat^2},0,\qthat^2,Q^2\) \log\frac{\qthat^2}{2(1+\cos\varphi)} +I_4.
\end{align}
The integral $I_4$, having no subtraction in it, can be computed as in Eq.~\eqref{eq:resCreg2},
using the delta function to fix $\varphi$.

Of course we can use the approach of
using the delta function to integrate over $\xi_2$ also for the resummed result.
In this case we find
\begin{align}
  \frac{Q^4\,\dd C_{\rm reg}}{\dd Q^2\dd y\dd\qt^2}&\(x,Q^2, y,\qt^2,\as, \frac{Q^2}{\muf^2}\) \nonumber\\
  &= \frac{\theta\(\frac{e^{-2|y|}}x-1-\qthat^2\)}{1+\qthat^2}
    \int_{\xi_0}^\infty \dd\xi_1 \int_0^{2\pi}\dd\varphi\, \frac{\theta\(\qthat^2-\xi_1(1-\cos^2\varphi)\)}{\sqrt{\qthat^2-\xi_1(1-\cos^2\varphi)}} U'_{\rm reg}\(\sqrt{x(1+\qthat^2)}e^{y},Q^2\xi_1,\muf^2\)\nonumber\\
  &\times \bigg[ \sqrt{\xi_2^+} U'_{\rm reg}\(\sqrt{x(1+\qthat^2)}e^{-y},Q^2\xi_2^+,\muf^2\) \frac{\dd{\cal C}}{\dd\varphi}\(\frac1{1+\qthat^2},\xi_1,\xi_2^+,Q^2,\varphi\) \theta\(\sqrt{\xi_2^+}-\sqrt{\xi_0}\) \nonumber\\
  &\quad +      \sqrt{\xi_2^-} U'_{\rm reg}\(\sqrt{x(1+\qthat^2)}e^{-y},Q^2\xi_2^-,\muf^2\) \frac{\dd{\cal C}}{\dd\varphi}\(\frac1{1+\qthat^2},\xi_1,\xi_2^-,Q^2,\varphi\) \theta\(\sqrt{\xi_2^-}-\sqrt{\xi_0}\) \bigg]
\end{align}
In fact, it is convenient to partition the integration region along the diagonal $\xi_1=\xi_2$, to get
\begin{align}
  \frac{Q^4\,\dd C_{\rm reg}}{\dd Q^2\dd y\dd\qt^2}\(x,Q^2, y,\qt^2,\as, \frac{Q^2}{\muf^2}\)
  &= \frac{\theta\(\frac{e^{-2|y|}}x-1-\qthat^2\)}{1+\qthat^2} \int_0^{2\pi}\dd\varphi \nonumber\\
  &\quad\times\[I_+\(x,Q^2,y,\qt^2,\as, \frac{Q^2}{\muf^2},\varphi\) + I_-\(x,Q^2,y,\qt^2,\as, \frac{Q^2}{\muf^2},\varphi\)\]\\
  I_+\(x,Q^2,y,\qt^2,\as, \frac{Q^2}{\muf^2},\varphi\)
  &=
    \int_{\xi_0}^{\frac{\qthat^2}{2(1+\cos\varphi)}} \dd\xi_1 \, \sqrt{\frac{\xi_2^+}{\qthat^2-\xi_1(1-\cos^2\varphi)}}
    \frac{\dd{\cal C}}{\dd\varphi}\(\frac1{1+\qthat^2},\xi_1,\xi_2^+,Q^2,\varphi\)\nonumber\\
  &\quad\times  U'_{\rm reg}\(\sqrt{x(1+\qthat^2)}e^{y},Q^2\xi_1,\muf^2\) U'_{\rm reg}\(\sqrt{x(1+\qthat^2)}e^{-y},Q^2\xi_2^+,\muf^2\) \\
  I_-\(x,Q^2,y,\qt^2,\as, \frac{Q^2}{\muf^2},\varphi\)
  &=
  I_+\(x,Q^2,-y,\qt^2,\as, \frac{Q^2}{\muf^2},-\varphi\).
\end{align}

\phantomsection
\addcontentsline{toc}{section}{References}

\bibliographystyle{jhep}
\bibliography{references}

\end{document}